\pdfoutput=1

\documentclass[11pt,a4paper]{article}
\usepackage[utf8]{inputenc}
\usepackage{fullpage}
\usepackage{hyperref}
\usepackage{amssymb}
\usepackage{amsthm}
\usepackage{graphicx}
\usepackage{color}
\usepackage{enumitem}
\usepackage{empheq}
\usepackage{cite}
\usepackage{cleveref}

\numberwithin{equation}{section}
\newcommand{\Li}[0]{\text{Li}}
\newcommand{\Gam}[0]{\Gamma}
\newcommand{\Dilog}[0]{\operatorname{Li}_2}
\newcommand{\Trilog}[0]{\operatorname{Li}_3}
\newcommand{\Disc}[0]{\operatorname{Disc}}
\newcommand{\Cut}[0]{\operatorname{Cut}}
\newcommand{\Mellin}[1]{\mathcal{M}_n\!\left[#1\right]}

\renewcommand{\Im}{\operatorname{Im}}

\newcommand{\pbar}{\bar{p}}
\newcommand{\nn}{\nonumber \\}

\newcommand{\e}{\epsilon}
\newcommand{\w}{\omega}

\def\eps{\epsilon}
\def\Ord{{\cal O}}

\def\d{{\rm d}}

\def\loopF{\mathcal{C}(\eps)}
\newcommand{\citeR}[1]{ref.~\cite{#1}}

\begin{document}

\begin{titlepage}
\hbox{NIKHEF/2016-012}
\vskip 30mm
\begin{center}
\Large{\sc{Unitarity methods for Mellin moments of Drell-Yan cross sections}}
\end{center}
\vskip 6mm
\begin{center}
Domenico Bonocore$^{1}$, Eric Laenen$^{1,2,3}$, Robbert Rietkerk$^{1,2}$\\[6mm]
\textit{$^1$Nikhef, Science Park 105, NL--1098 XG Amsterdam, The Netherlands}\\[1mm]
\textit{$^2$ITFA, University of Amsterdam, Science Park 904, Amsterdam, The Netherlands}\\[1mm]
\textit{$^3$ITF, Utrecht University, Leuvenlaan 4, Utrecht, The Netherlands}\\[1mm]
\end{center}
\vspace{0.5cm}

\bibliographystyle{utphys}

\begin{abstract}
We develop a method for computing Mellin moments
of single inclusive cross sections such as Drell-Yan production
directly from forward scattering diagrams, by invoking unitarity in the form of cutting
equations.  We provide a diagram-independent prescription for
eliminating contributions from unwanted cuts at the level
of an expansion in the reciprocal $\omega=1/z$ variable.
The modified sum over powers of $\omega$ produces the result 
from physical cuts only, with the $n$th coefficient precisely equal
to the $n$th Mellin moment of the cross section. 
We demonstrate and validate our method for representative one- and two-loop diagrams.
\end{abstract}

\vskip -1mm
\noindent 
\end{titlepage}

\pagebreak

\section{Introduction}
\label{sec:introduction}

The efficient computation for higher order QCD corrections for scattering processes
has been a mainstay of research in theoretical particle physics for decades, 
as this directly impacts the potential for discovery at colliders: the more precisely signal and background are
computed, the more significant comparison of theory with data can be. 
In recent years a notable increase in computational capacity has taken place, spurred largely
by the development of unitarity techniques for computing scattering amplitudes. The ability
to determine a scattering amplitude from its poles and branch cuts \cite{'tHooft:1973pz,Bern:1996je,Berger:2009zb} 
has been a watershed in these efforts, especially for one-loop high-multiplicity processes.

Unitarity methods for few external legs but at higher loop have proven to be highly valuable as well.
Reverse unitarity techniques have been important in relating real emission amplitudes
to virtual ones \cite{Anastasiou:2002yz} for two and three-loop calculations.
An inspiration for the present paper is the body of work 
on the computation of the 2- and 3-loop splitting functions and
deep-inelastic scattering (DIS)
Wilson coefficients \cite{Vermaseren:2005qc,Vogt:2004mw,Moch:2004pa,Moch:1999ebq,Mitov:2005ps,Mitov:2006wy}.
Computing the DIS structure functions by moments has been a very
succesful approach for massless partons
\cite{Larin:1993vu,Larin:1996wd,Retey:2000nq,Kazakov:1987jk,Moch:1999ebq,Vermaseren:2005qc},
and also for heavy quarks \cite{Bierenbaum:2009mv,Blumlein:2009rg,Behring:2014eya,Blumlein:2012vq,Ablinger:2010ty,Ablinger:2014uka,Ablinger:2012qm,Ablinger:2014yaa,Ablinger:2015tua}.
For the inclusive DIS structure functions it is possible to use the optical theorem 
to compute all the cuts of the forward scattering process, and directly extract
the Mellin moments of the coefficients. 
In essence, from unitarity considerations, one may expand the forward scattering 
amplitude in reciprocal powers of the Bjorken scaling variable $x$, the coefficient
at order $n$ then being the $n$th Mellin moment of the coefficient function.

In this paper we aim to generalise this method to a single-inclusive cross section,
specifically the Drell-Yan cross section, which is then prototypical for other processes such
as Higgs production in the large top mass limit, 
for which remarkable results for high-order corrections to Higgs
have recently been achieved
\cite{Anastasiou:2015ema,Anastasiou:2014lda,Anastasiou:2016cez,Catani:2014uta,Bonvini:2014jma,Ahmed:2014cla}.
The optical theorem
does not directly apply to the Drell-Yan cross section,
not being a fully inclusive observable.
We show in this paper  that it is nevertheless possible to compute the Mellin moments
of the Drell-Yan cross section directly from forward diagrams, using unitarity and an expansion in reciprocal powers of 
$z$. The key aspect of our method is the efficient subtraction
of unwanted cuts, through complex-valued shifts of the moment variable
$n$ and through a replacement rule for harmonic sums.

The paper is organised as follows. In section 2 we review the relevant aspects of
unitarity and the optical theorem, and their role in DIS. 
In section 3 we treat the one-loop corrections in ``scalar'' Drell-Yan correction with our new method in
some detail, highlighting key features. In section 4 we test our
method at two loops, showing how our methods work for representative two-loop forward scattering
scalar diagrams. Here we show explicitly how remove contributions
from unphysical cuts, such that those from physical cuts are unaltered. 
We conclude with a summary and some remarks towards further development.

\section{Forward amplitudes and unitarity}
\label{sec:forward-amplitudes-and-unitarity}
In this section we outline the general ideas of the paper,
postponing technical details to the following sections, where 
one and two loop examples will be discussed.

We start in \cref{sec:DIS-optical-theorem} reviewing the essential points that make the optical
theorem successful for DIS.
Then, in \cref{sec:analytic-structure} we move to the Drell-Yan case, stressing the differences 
that make a generalization of the DIS method highly non-trivial.
Among these, the most problematic one is the presence
of unphysical cuts, absent in DIS, that need to be removed from the discontinuity of the forward amplitude. 
Therefore in \cref{sec:classification-of-cuts} we classify all unphysical cuts,
showing that most of them either vanish or are easily treated.
For the remaining unphysical cuts, we outline a solution in \cref{sec:extract-correct-cut}, 
referring the reader to \cref{sec:DY-one-loop} and \cref{sec:DY-two-loop} for more technical explanations.

\subsection{DIS and the optical theorem} 
\label{sec:DIS-optical-theorem}

Let us first review the role of the unitarity, in the form of the
optical theorem, in deep-inelastic scattering through off-shell photon
exchange. Our exposition follows largely that of \citeR{Sterman:1994ce}.
It is well-known that the fully inclusive cross section for this
process, $e(l)+P(p) \longrightarrow e(l') + X$, can be written in the
form
\begin{equation}
  \label{eq:1}
  d\sigma = \frac{1}{2s}\frac{1}{Q^4} L^{\mu\nu}(l,l') W_{\mu\nu}(p,q)
  \frac{d^3l'}{|l'|}\,,
\end{equation}
with $q= l-l'$, and with $W_{\mu\nu}$ ($L_{\mu\nu}$) the hadronic
(leptonic) tensor. They are defined as
\begin{align}
  & L_{\mu\nu}(l,l') = \frac{e^2}{8\pi^2}\left( l_\mu l'_\nu + l_\nu l'_\mu - \eta_{\mu\nu}
  l\cdot l'\right)\,, \nn
  & W_{\mu\nu}(p,q) = \frac{1}{8\pi}\sum_n \langle P(p) |
  J^\dagger_\mu (0) | n\rangle (2\pi)^4 \delta^{(4)} (p_n-p-q) \langle
  n | J_\nu(0) | P(p) \rangle\,,
   \label{eq:4}
\end{align}
where implicit spin quantum numbers in the external states are 
summed over. Note that the sum over final states $|n\rangle$ is fully
inclusive in terms of QCD as both explicit momenta $p$ and $q$ are
incoming.
Current conversation and parity invariance in both indices then imply
the structure
\begin{equation}
  \label{eq:3}
  W_{\mu\nu}(p,q) = - \left(\eta_{\mu\nu} - \frac{q_\mu q_\nu}{q^2}
  \right)  W_1(x, Q^2)
+ \left(p_\mu - \frac{q_\mu p\cdot q}{q^2}  \right) 
  \left(p_\nu - \frac{q_\nu p\cdot q}{q^2}  \right)  W_2(x, Q^2)\,,
\end{equation}
so that the structure of the proton is encoded into two scalar
functions that depend on the variables
\begin{equation}
  \label{eq:7}
  Q^2 = -q^2, \qquad x = \frac{Q^2}{2p\cdot q}\,.
\end{equation}

The optical theorem applies to the hadronic tensor $W_{\mu\nu}$, since the sum in \cref{eq:4} is fully inclusive so that we can write
\begin{equation}
  \label{eq:5}
  W_{\mu\nu} (p,q)= 2\Im T_{\mu\nu}(p,q)  \,,
\end{equation}
where $T_{\mu\nu}$ is the forward Compton amplitude
$\gamma^*(q)+P(p)\longrightarrow  \gamma^*(q)+P(p)$, having
the same tensor structure as in \cref{eq:3} but now in terms of 
the scalar functions 
\begin{equation}
  \label{eq:6}
  T_i\left(\w=\frac{1}{x},q^2\right), \quad i =1,2\,.
\end{equation}
Note that we have chosen to indicate the functional dependence in terms of the \emph{reciprocal}
$x$ variable, for reasons we discuss below.
The functions $W_i$ and $T_i$ both have
cuts starting at branch points at $x=\pm 1$, 
corresponding to the kinematical conditions for normal thresholds, $(p\pm q)^2>0$.
For the $W_i(x,Q^2)$ functions, the cut then
runs from $x=-1$ to $x=1$. For the $T_i(\w,Q^2)$ functions, consequently, the cuts lie along the
$\w$-intervals $(-\infty,-1]$ and $[1,\infty)$. 
These are also the only cuts, and we have in general for the $T_i$
\begin{equation}
  \label{eq:8}
  T_i(-\w,Q^2) =   T_i(\w,Q^2) \,.
\end{equation}

\paragraph{}

\begin{figure}[t]
\begin{center}
\includegraphics{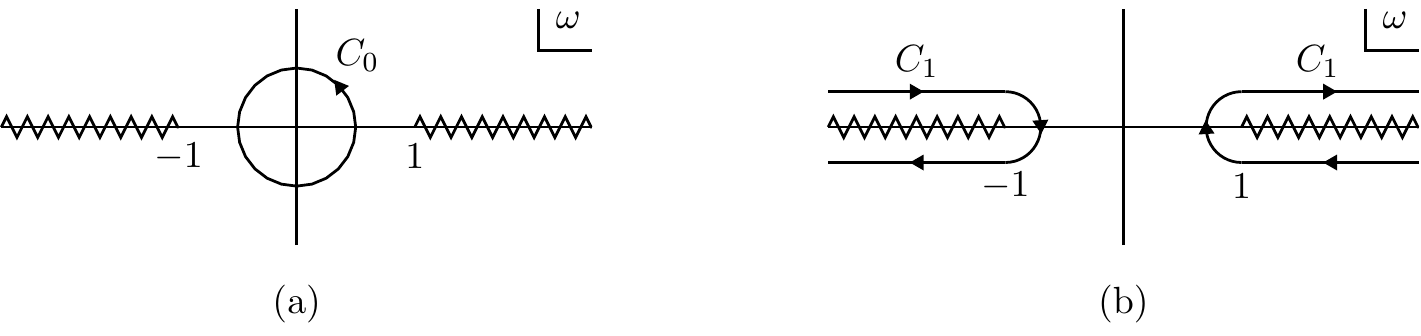}
\caption{
Branch cut structure of $T_i(\w,Q^2)$ 
with the two contours used for $T^{(n)}_i(Q^2)$.
On the left the contour $C_0$  wraps around the origin,
while on the right the contour $C_1$ encloses the two branch cuts.
Note that the combination $\w^{-n-1}T(\w)$ 
of \cref{eq:9} has an
additional pole at the origin. 
}
\label{fig:ImT_C1Contour}
\end{center}
\end{figure}

One can now compute Mellin moments of the
$W_i(x,Q^2)$ functions by expanding the $T_i(\w,Q^2)$
amplitudes.  The $n$th derivative of $T_i$ at $\w=0$ may be
rewritten by Cauchy's theorem in terms of the contour in \cref{fig:ImT_C1Contour}a
\begin{equation}
  \label{eq:9}
T_i^{(n)}(Q^2) 
\equiv \frac{1}{n!}\frac{d^n  T_i(\w,Q^2)}{d\w^n} \bigg{|}_{\w=0} 
= \oint_{C_0} \frac{d\w}{2\pi i} \w^{-n-1} T_i(\w,Q^2)\,.
\end{equation}
The contour $C_0$ may be deformed into the contour $C_1$ shown in 
\cref{fig:ImT_C1Contour}b. Then, 
using \cref{eq:8}, 
we have
\begin{align}
T_i^{(n)}(Q^2)
&=  \frac{\left(1+(-1)^{n}\right)}{2\pi i} \int_1^{\infty}d\w\,
 \w^{-n-1}\, \underset{\w}{\Disc} \,T_i(\w,Q^2)~ ,
\label{eq:TnfromDisc}
\end{align}
where the discontinuity of a function in 
the variable $x$ is defined in general by 
\begin{align}
\underset{x}{\Disc} f(x) = \lim_{\eta \rightarrow 0} \big( f(x+i\eta) - f(x-i\eta) \big) ~.
\end{align}
The presence of the factor $\left(1+(-1)^{n}\right)$ in \cref{eq:TnfromDisc} implies that odd series coefficients vanish.
For $n$ even, instead, using the optical theorem in the form of \cref{eq:5}, and changing integration variables to $x=1/\w$, we get 
\begin{equation}
  \label{eq:10}
  T_i^{(n)}(Q^2) = \frac{1}{\pi} \int_0^1 dx \, x^{n-1} W_i(x,Q^2) =
\frac{1}{\pi} 
\Mellin{W_i(Q^2)}~,
\end{equation}
where the second equality 
defines the Mellin transform ${\cal M}_n$. Thus, indeed, the expansion of the forward
scattering amplitude in $\w$ yields the Mellin moments of the
cross section.

A few remarks are in order. This way of using the optical theorem, computing
directly the Mellin moments of the DIS structure functions by
expansion in $1/x$, has been marvellously successful for 2- and 3-loop
calculations for DIS \cite{Vermaseren:2005qc,Moch:2004pa,Vogt:2004mw}. 
Translating back to momentum space 
is readily done, and produces known combinations of functions (Harmonic Polylogarithms
(HPL's) \cite{Remiddi:1999ew}).  The presence of the branch points at
$\w=\pm 1$ and the analytical behaviour of the $T_i(\w,Q^2)$
near $\w=0$ is very helpful towards the consistency and also
practicality of the method. 

The question whether the DIS method can be generalized to semi-inclusive cross sections such as Drell-Yan and Higgs production will be addressed in the next sections, and is indeed the central issue of the present paper.

\subsection{Analytical structure of one-particle inclusive processes}
\label{sec:analytic-structure}

We consider the inclusive production of an
 electroweak boson $V$ 
of invariant mass $Q^2$
by
quark-antiquark annihilation 
\begin{align}
  q(p)+\bar q(\bar p) \longrightarrow  V +X ~,
\end{align}
where $X$ represents any partonic contribution to the final state and the vector
boson $V$ may be an off-shell photon $\gamma^*$, or an on-shell
$W^{\pm}$ or $Z$ boson. 
As such, this process is described by two scales only, 
the mass $Q^2$ and  
the squared partonic center-of-mass energy~$s$, from which 
it is possible to define the dimensionless variables
\begin{align}
\w=\frac{1}{z}, \quad ~z=\frac{Q^2}{s}~,
\end{align}
where $z$ is the variable analogous to the Bjorken variable $x$ for DIS.
In the following we will
focus on the case of an off-shell photon,
referring  to this process as Drell-Yan.
Apart from interesting in its own right, this process is prototypical
for many other partonic processes relevant at hadron colliders, especially
Higgs boson production via gluon fusion. Differences with
the Drell-Yan case resides only in numerator factors which go along for the ride in our method.
Before discussing how the generalisation of the DIS formalism to the
Drell-Yan case can be set up, let us review those similarities and
differences between the two processes that are relevant for our
purposes.

\paragraph{}

At face value, the differences seem not very large.  
Focusing on the partonic part, the set of
diagrams for the Drell-Yan process can be obtained from the DIS ones, by crossing the exchanged off-shell
photon to the final state, and the outgoing quark to the initial
state.  However, this crossing has significant consequences.  
First, the off-shellness of the photon becomes time-like and can be 
effectively regarded as a mass. Therefore the forward amplitude
$  q(p)+\bar q(\bar p)\longrightarrow  q(p)+\bar q(\bar p)$
will contain a massive propagator. 
Then, most importantly, the vector boson must be present in 
the final state. Hence, the process is not fully inclusive like DIS, but
only single-particle inclusive,
so that the optical theorem, as the simplest
realization of unitarity, cannot be used and the cross section is not
given by the imaginary part of the full forward amplitude.

A further complication arises when moving to Mellin space.  Looking
at the analytical structure of the DIS forward amplitude, branch cuts in
the $\w$-plane are at $(-\infty,-1]$ and $[1,\infty)$.  Due to the
symmetry in \cref{eq:8} of the forward amplitude, one may
consider only the cut along the positive real axis, which can be eventually
converted to a Mellin transform, as discussed in the previous subsection. 
In the Drell-Yan case instead  
the forward amplitude generally will have more 
branch cuts, 
in particular also along $(-\infty,0]$ and $[0,\infty)$
and no symmetry relates
the forward amplitudes with opposite value of $\w$.
A new strategy is needed if we want to extract the series
coefficients of the forward amplitude
 expanding it around the branch point $\w=0$.

These considerations suggest that extending the DIS techniques
for directly computing Mellin moments
to the Drell-Yan case is not straightforward.
However, we shall see that  it is possible when using unitarity cuts.
Let us discuss the key aspects of this in somewhat more detail. 

\paragraph{}

The optical theorem relates a cross section to the imaginary part of the relevant forward amplitude.  
At the same time, this imaginary part is, by the Cutkosky rules \cite{Cutkosky:1960sp}, 
equal to the sum over all cuts of the amplitude. 
For a fully inclusive processes like DIS,
 these cuts correspond to the phase space integration
 of the squared matrix elements of the process. 
For Drell-Yan instead this is not the case, as cuts that do not cut
the massive photon are not to be included for the cross-section.  
However, the use of unitarity cuts is considerably more general,
and holds on diagram-by-diagram basis.
In general, branch cut discontinuities in different channels 
correspond to different cuts of the diagram \cite{Veltman:1994wz,Veltman:1963th} 
\begin{align}
\Disc \mathcal{F} = \sum_{k} \Cut_k \mathcal{F} ~,
\label{eq:general_cutting_equation}
\end{align}
for any Feynman diagram $\mathcal{F}$.
Our approach exploits this fact fully when ${\cal F}$ is a forward scattering
diagram.

Our goal is to compute (moments of) the cross section from the forward
amplitude.  The cross section can be reconstructed from the
discontinuity of the forward amplitude across the physical branch cut.
In general an amplitude has discontinuities around unphysical branch
cuts as well, and these must be subtracted.  This does not seem a very
efficient procedure, as it apparently requires one to compute unphysical-cut 
diagrams nonetheless. Moreover, the unphysical cuts may be even more complicated than
the physical cuts.  However, as we will see in the following
sections,  one can modify the
analytic structure of the forward amplitude such that its
discontinuity is given by the sum of physical cuts only.  In
particular, we will see that, after moving to Mellin space, it is
possible to automatically select the physical cut without the need to
subtract (and compute) the unphysical cuts.

Let us first review the classification of
the cuts appearing in the forward amplitude of the Drell-Yan process. 

\subsection{Classification of Drell-Yan cuts}
\label{sec:classification-of-cuts}

\begin{figure}[t]
\begin{center}
\includegraphics{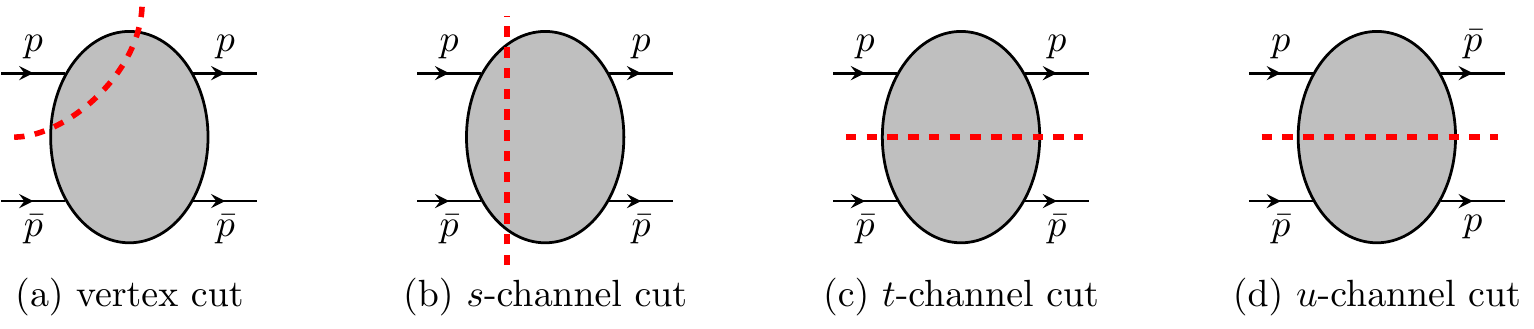}
\caption{
Generic cuts of forward aplitudes with two initial massless particles with momenta $p$ and $\pbar$. The cuts of type (a) and (c) vanish by the general cutting rules. Note that the $u$-channel cut differs from the $t$-channel cut by interchange of the two outgoing momenta.}
\label{fig:generic-cuts-of-foward}
\end{center}
\end{figure}

The set of diagrams we would like to discuss are those required in a
NNLO calculation, though
many of the features will be valid also at higher order.
To set up our classification, 
we regard forward diagrams as amplitudes that may depend on different channels, and therefore can be cut 
in all possible ways, as long as the diagram is cut into two (connected) subgraphs.
In this regard, we distinguish four different classes of cuts,
depicted in \cref{fig:generic-cuts-of-foward} and denote them as 
vertex, $s$-channel, $t$-channel and $u$-channel cuts.
Of course, for a forward amplitude and with on-shell external lines
the only possible invariant is $s$, but the nomenclature will be useful,
and is based on the case when final momenta are different from the
initial ones, such that also the $t$ and $u$ channels would be open.

The $s$-channel cuts are the only ones that can be interpreted as 
a phase space integration of squared matrix elements. 
Among these, \emph{physical} cuts, i.e. those that contribute to the cross-section,
are only those $s$-channel cuts that pass through the massive photon,
and we thus call \emph{massive} $s$-channel cuts. 
More generally, we call massive (massless) every cut that does (does not) cut the massive boson propagator.  
At first it seems that the number of unphysical cuts might grow dramatically 
with the order of the computation, making it difficult to control them.
However, a number of simplifications are possible, making some of
these cuts give a vanishing contribution.

The vertex cut in \cref{fig:generic-cuts-of-foward}a vanishes, because it measures the discontinuity of the forward amplitude in the $p^2$-channel. 
But this discontinuity is zero, because the forward amplitude does not actually depend on this variable due to the on-shell condition $p^2=0$.
The same holds when any other of the four vertices is cut.
Furthermore, by the same token, the $t$-channel cut in \cref{fig:generic-cuts-of-foward}c vanishes as well.
This leaves only the $s$- and $u$-channel cuts to be considered:
the massless $s$-channel cuts and the (massive and massless) $u$-channel cuts.
In \cref{sec:unphysical-cuts-two-loops} we will treat these unphysical cut diagrams in greater detail and in specific examples.  
In the following subsection we first review the general ideas how to deal with these cuts.

\subsection{Extracting the physical cuts from the forward amplitude}
\label{sec:extract-correct-cut}

At this point we make an important observation.  The forward amplitude
carries more information than needed; indeed we are only interested in
its imaginary part.  We have the freedom to modify the amplitude, as
long as the branch cut structure remains the same.  For instance,
adding a constant or even an analytic function will not affect the
physical information one wishes to extract from its cuts.  This
consideration leads us to disregard lower order Mellin moments. 
Indeed, assuming that the forward amplitude can be expanded
around $\w=0$ as
\begin{equation}
\label{eq:sumn0}
  f(\w)=\sum_{n=n_0}^{\infty} c_n \, \w^n~, \qquad n_0\geq 0
\end{equation}
its series coefficients $c_n$ will be defined only for $n \geq n_0$.
However, any shift in $n_0$ making the sum start from a new positive
integer is equivalent to adding to the original $f(\w)$ simply polynomials
in $\w$, which does not affect the branch cut structure.  Therefore, we
conclude that no physical information is carried in the lower bound of
the sum,
and henceforth we shall omit it
in series expansions except where necessary.
We can even take a further step in this line of reasoning.  Since we
are interested in extracting the discontinuity of the forward
amplitude only across the \emph{physical} branch cut, we have the
freedom to redefine the forward amplitude, modifying also its branch
cut structure, as long as this leaves the physical branch cut
unaltered. Also, poles in $\w$ can be removed from (\ref{eq:sumn0}) because such poles do not correspond to physical cuts.
These steps form the essence of the strategy we shall
implement to deal with the unphysical cuts.

We start with the first two types of cuts presented in \cref{sec:classification-of-cuts}: massless $s$-channel cuts and massless $u$-channel cuts. 
These classes of cut diagrams contain
factors $\w^{\e}$, where $\eps = \frac{4-d}{2}$ is the dimensional regulator. Hence, they belong to branch cuts starting at $\w=0$. As such, those are the cuts that prevent the forward amplitude to be written as a power series in $\w$ around $\w=0$.
We introduce a \emph{shifting procedure}, through which it will be possible to define a new function $\widetilde f(\w)$ with no branch point at $\w=0$.  
This will be necessary already at one loop and will be further discussed in \cref{sec:shifting-procedure}.

The most difficult kind of unphysical cut appears only from two loops: the massive $u$-channel cut, because it corresponds to a branch cut in the forward amplitude starting at $\w = -1$.
We shall go beyond the simple shifting procedure and subtract the contributions from this cut directly in Mellin moment space, following the extending reasoning above. 
We are able to compose a dictionary of \emph{replacements} 
for harmonic sums, which may be applied to any diagram.  
This procedure will be first discussed in \cref{sec:approach-two-loop-amplitudes} and is applied to a two-loop crossed box in \cref{sec:two-loop-crossed-box}.

After all unphysical cuts are removed, one can 
repeat the procedure carried out for DIS in \cref{fig:ImT_C1Contour}.
This time only the branch cut $[1,\infty)$ is present and
the generalization of \cref{eq:10} (for both even and odd Mellin moments) reads
\begin{align}
\mathcal F_{\text{phys}}^{(n)}=\frac{1}{2\pi i} \, \Mellin{\Cut_{\text{phys}}\mathcal F}~,
\label{eq:cutting_Mellin}
\end{align}
where $\mathcal F_{\text{phys}}^{(n)}$ are the series coefficients of the
forward amplitude modified such that it contains only the physical cut. 
Let us now turn to a more explicit illustration of these methods at
one-loop.

\section{Drell-Yan at one loop}
\label{sec:DY-one-loop}

\begin{figure}[t]
\begin{center}
\includegraphics{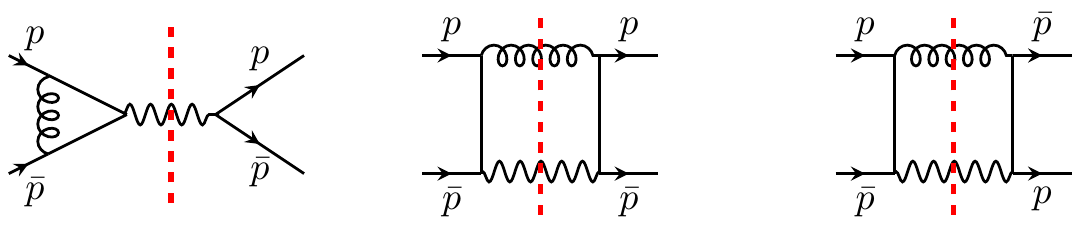}
\caption{Diagrams needed for the one-loop DY cross sections. 
Diagrams obtained by complex conjugation or 
exchanging $p \leftrightarrow \bar p$  
are omitted. Arrows on the lines indicate momentum flow.}
\label{fig:one-loop-full-QCD}
\end{center}
\end{figure}

Here we develop the concepts from \cref{sec:forward-amplitudes-and-unitarity} 
for the one-loop Drell-Yan cross section.  The standard
approach requires the evaluation of the matrix elements for real and
virtual corrections, and the subsequent integration over the phase
space.  This requires the computation of three phase space integrals,
represented in \cref{fig:one-loop-full-QCD}, which in the language of
\cref{sec:forward-amplitudes-and-unitarity} are massive $s$-channel
cuts.
In order to minimise technical complications we will henceforth omit numerator factors
in those diagrams, i.e. we restrict to the scalar case, as these are anyway
irrelevant for illustrating our method. This
is motivated by the fact that the analytical structure
of Feynman integrals in QCD arises in essence from denominators in diagrams.%
\footnote{We have of course checked that the well-known one-loop results
for Drell-Yan \cite{Altarelli:1979ub} are recovered after restoring the numerators.}

In the next three subsections the calculation of the scalar equivalent
of the diagrams in \cref{fig:one-loop-full-QCD} is presented,
verifying the cutting equation in Mellin space in
the form of \cref{eq:cutting_Mellin}.  
In this process we encounter
two relatively harmless types of unphysical cuts and we show how to
deal with them as discussed in \cref{sec:forward-amplitudes-and-unitarity}.  
In particular, we devote one subsection to the notion
of the shifting procedure, which is needed to 
remove both kinds of cuts.
Given the simplicity of those one-loop calculations, the computation
is actually performed to all orders in $\eps$. However, extracting the series
coefficients of the $\w$ expansion exactly in $\eps$ 
is not feasible at higher loops.
Therefore, in the last subsection section we illustrate how to compute
the Mellin moments using IBP identities, by the example
of the one-loop box diagram.

\subsection{Triangle diagram}
\label{sec:one-loop-triangle}

\begin{figure}[t]
\begin{center}
\includegraphics{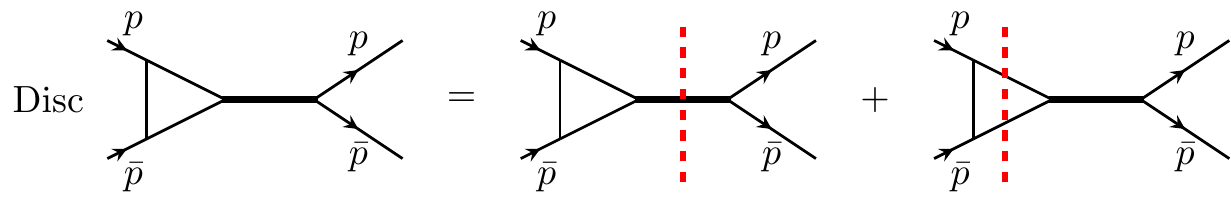}
\caption{
Cutting equation for the one-loop triangle diagram. The right-hand side features both $\Cut_1 T$, a physical
massive $s$-channel cut, and $\Cut_2 T$, an unphysical massless $s$-channel cut. 
}
\label{fig:one-loop-triangle-cutting-equation}
\end{center}
\end{figure}

The simplest of the three diagrams that contribute to the one-loop Drell-Yan forward amplitude involves a triangle loop. 
The cutting equation for this graph is schematically depicted in \cref{fig:one-loop-triangle-cutting-equation}, showing both a physical and an unphysical cut. 
It appears that in order to compute the physical $\Cut_1 T$ contribution one would need to `subtract' the unphysical $\Cut_2 T$ contribution from the full discontinuity $\Disc T$.

To verify  explicitly the
presence of an unphysical cut,
 let us compute both the forward amplitude and the 
physical cut diagram.
The forward amplitude $T$ reads
\begin{align}
T&=\loopF \,Q^{2\e}\, \frac{s^2}{s-Q^2}\,\int \frac{\d^{4-2\eps}k}
{k^2\,(k-p)^2\,(k-p-\pbar)^2} \nonumber \\
&=\frac{1}{\eps^2}\,
(-\w)^{-\eps}\, \frac{\w}{\w-1} ~,
\label{eq:T_fwd}
\end{align}
where the prefactor coming from the loop integration is given by
\begin{align}
\loopF = \frac{1}{i \pi^{2-\eps} r_\Gam} ~,~
r_\Gam = \frac{\Gam(1-\eps)^2\,\Gam(1+\eps)}{\Gam(1-2\eps)}\,.
\end{align}
and we have set $\mu^2=Q^2$.
We rescale forward diagrams $\cal F$ by their mass dimension $s^{\text{dim}[\cal F]}$, such that they become dimensionless.
Applying the Cutkosky cutting rule, the physical cut reads
\begin{align}
\Cut_1 T&=\loopF \,Q^{2\e}\,  (-2\pi i) \, s^2\, \delta(s-Q^2) \,\int \frac{\d^{4-2\eps}k}
{k^2\,(k-p)^2\,(k-p-\pbar)^2} \nonumber \\
&=-\frac{2\pi i}{\eps^2}\,
(-z)^{\eps}\, \delta(1-z) ~,
\end{align}
where we expressed the result as function of $z=1/\w$.
Clearly, the discontinuity of the forward amplitude is not given 
by the physical cut alone, as can be seen by the presence
of $(-\w)^{-\e}$ in the forward amplitude.%
\footnote{The importance of keeping track of such factors $z^{-\eps}$ has recently also been analysed in the context of $e^+e^-$ annihilation at two loops with differential equations \cite{Gituliar:2015iyq}.} 
Indeed, upon expanding this factor in $\e$,
it is evident that $T$ has a branch cut for $\w<0$, 
whereas $\Cut_1 T$ is different from zero only at $\w=1$.
Therefore, to recover the physical cut diagram, we should compute
also the unphysical $\Cut_2 T$.
As we shall see, this can be bypassed in Mellin space.

We start computing the Mellin moments of the physical cut
\begin{align}
\Mellin{\Cut_1 T} 
\,= \,- \frac{2\pi i}{\eps^2}\,\int_0^1 z^{n-1}(-z)^{\eps}\, \delta(1-z) 
\,=\,- \frac{2\pi i}{\e^2}\, e^{- i \pi \e}~,
\label{eq:triangle_Mellin}
\end{align}
where the phase $e^{- i \pi \e}$ is due to the minus sign in $(-z)^{\e}$
and can be fixed by keeping track of the Feynman $i\eta$ in the propagators.
Then, we would like to
compare \cref{eq:triangle_Mellin}
with  the series coefficients $c_n$
of the forward amplitude expanded 
in powers of $\w$.
However, as can be seen in \cref{eq:T_fwd}, this expansion cannot be perfomed,
since $T$ contains a non-integer power of $\w$. This is of course expected: the forward amplitude has a branch point at $\w=0$ and therefore cannot be expanded around that point. However, the structure of this branch cut starting from the origin is simply given by $(-\w)^{-\e}$. 
Therefore, we apply the following procedure. 
We write a series representation for the other factors, to get
\begin{align}
T &=-\frac{e^{-i \pi \eps}}{\eps^2}\,
 \sum_{n=1}^{\infty}\w^{n-\e} ~.
\label{eq:T_fwd2}
\end{align}
Then, we shift $n \to n+\e$ in the summand,
leaving the lower bound of the sum unaltered.
This defines a \emph{new} function $\widetilde T$ 
\begin{align}
\widetilde T= 
-\frac{e^{-i \pi \eps}}{\eps^2}\,
 \sum_{n=1}^{\infty}\w^{n}\,=\,\frac{e^{-i \pi \eps}}{\eps^2}\,\frac{\w}{\w-1} ~,
\label{eq:T_fwd3}
\end{align}
with no branch cut
from the origin, but with the physical pole in $\w=1$ still present. 
This procedure will be also used in 
\cref{sec:one-loop-crossed-box} for the crossed box diagram,
where a non-trivial $n$-dependence
of the summand will make the shift less trivial. 
Thus, writing $\widetilde T = \sum_n \widetilde c_n \w^{n}$, we find
\begin{align}
\widetilde c_n=-\frac{e^{-i \pi \eps}}{\eps^2}~.
\label{eq:triangle_c_tilde}
\end{align}
Comparing \cref{eq:triangle_Mellin} and \cref{eq:triangle_c_tilde}
we get
\begin{align}
\widetilde c_n = \frac{1}{2\pi i}\,\Mellin{\Cut_1 T}~,
\end{align}
which verifies the cutting equation in Mellin space in the form 
of \cref{eq:cutting_Mellin}.

The triangle diagram is the easiest example that exhibits an unphysical cut and where it is possible to test our Mellin space approach.
In the next subsection we discuss an example with non-trivial $n$-dependence.

\subsection{Box diagram}
\label{sec:one-loop-normal-box}

\begin{figure}[ht]
\begin{center}
\includegraphics{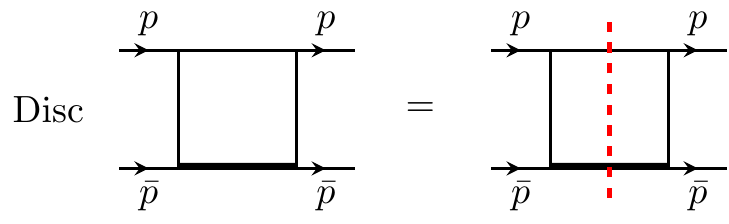}
\caption{Cutting equation for the one-loop box diagram.}
\label{fig:one-loop-box-cutting-equation}
\end{center}
\end{figure}

For the box diagram $B_1$ all unphysical cuts vanish at the outset.
The only cut, shown in \cref{fig:one-loop-box-cutting-equation}, is the physical massive $s$-channel cut, which we call $\Cut_1 B_1$.

As we did for the triangle diagram, let us compute
explicitly 
the forward amplitude $B_1$ and the cut diagram $\Cut_1 B_1$.
From a direct computation we have
\begin{align}
B_1&=\loopF\,Q^{2\e}\,s^2\,\int\frac{\d^{4-2\eps}k}
{((k+\pbar)^2)^2\,(k+p+\pbar)^2\,(k^2-Q^2)} 
\nn&=
\frac{\Gam(1-2\e)}
{\e\,\Gam(1-\e)\,\Gam(2-\e)} \,\w^2\,
~{}_2 F_1(1,2+\eps;2-\eps;\w)
\label{eq:B1_fwd}~,
\end{align}
where we expressed the result as a function of $\w=s/Q^2$.
The cut diagram $\Cut_1 B_1$ is easily computed as well.
It is defined as
\begin{align}
\Cut_1 B_1&=\loopF\,Q^{2\e}\,(-2\pi i)^2\,\int\d^{4-2\eps}k\,
\frac{\delta^+(k^2)\,\delta^+((p+\pbar-k)^2-Q^2)}
{(k-p)^2\,(k-p)^2} ~.
\end{align}
As for all massive $s$-channel cuts, this integral is non-vanishing when $s>Q^2$, and we can perform the computation in the centre-of-mass frame, where
\begin{align}
p = \frac{\sqrt{s}}{2} (1,1,0,0) ~,~~
\pbar = \frac{\sqrt{s}}{2} (1,-1,0,0) ~,~~
k = k^0 (1,\cos\theta,\sin\theta,0) ~.
\end{align}
Computing the integral in this frame yields
\begin{align}
\Cut_1 B_1&=
- 4 i \pi \,
z^{\e}\,(1-z)^{-1-2\e}
\frac{\theta(z)\,\theta(1-z)}{\Gam(2+\eps)\,\Gam(1-\eps)}~,
\label{eq:B1_cut}
\end{align}
where we expressed the result as a function of $z=1/\w$.
Note that both results (\ref{eq:B1_fwd}) and (\ref{eq:B1_cut}) are valid to all orders in $\eps$. 

We can now verify the cutting equation in Mellin space.
Specifically,  the forward amplitude may be written as a series representation $B_1 =\sum_{n} c_n\, \w^n$,
where
\begin{align}
c_n=
\frac{\Gam(1-2\eps)}{\eps\,\Gam(1-\eps)\,\Gam(2+\eps)} \,
 \frac{\Gam(n+\eps)}{\Gam(n-\eps)}~, 
\label{eq:B_1_series}
\end{align}
while the Mellin transform of the cut diagram can be trivially computed 
and reads
\begin{align}
\Mellin{\Cut_1 B_1} &=
- 4 \pi i \,
\frac{\Gam(-2\eps)}{\Gam(2+\eps)\,\Gam(1-\eps)} \,
\frac{\Gam(n+\eps)}{\Gam(n-\eps)}  ~.
\label{eq:B_1_mellin}
\end{align}
Comparing \cref{eq:B_1_series} and \cref{eq:B_1_mellin} we find
\begin{align}
c_n = \frac{1}{2 \pi i} \Mellin{\Cut_1 B_1}
\end{align}
which verifies the cutting equation in Mellin space in the form of \cref{eq:cutting_Mellin}.

This example has shown the cutting equation 
in Mellin space with a non-trivial $n$-dependence.
To increase further the complexity, 
in the next section we will compute the
 crossed-box diagram, which has both non-vanishing unphysical cuts and non-constant moments.

\subsection{Crossed-box diagram}
\label{sec:one-loop-crossed-box}

\begin{figure}[ht]
\begin{center}
\includegraphics{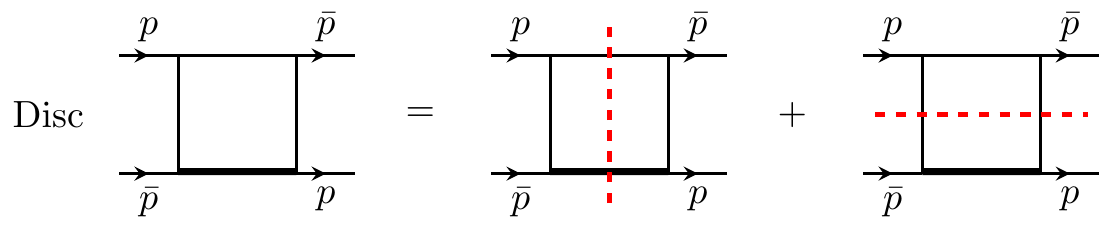}
\caption{Cutting equation for the one-loop crossed-box diagram $B_2$, featuring on the right-hand side both the physical 
$s$-channel cut ($\Cut_1B_2$) and the unphysical $u$-channel cut ($\Cut_2 B_2$).}
\label{fig:one-loop-crossed-box-cutting-equation}
\end{center}
\end{figure}

The crossed box $B_2$ is the last contribution to the one-loop Drell-Yan forward amplitude.
It results from the normal box $B_1$ by interchanging the two final state momenta.
This has the consequence that the $u$-channel cut does not vanish so that the discontinuity of the forward consists of two cuts, a physical and an unphysical cut,
as shown in \cref{fig:one-loop-crossed-box-cutting-equation}.

As we did for the other two examples,
we compute the forward amplitude and the cut diagrams. The former it is defined as
\begin{align}
B_2&=\loopF \,Q^{2\e}\,s^2\,\int\frac{\d^{4-2\eps}k}{(k+p)^2(k+\pbar)^2(k+p+\pbar)^2(k^2-Q^2)} ~.
\end{align}
Again, applying standard techniques with Feynman parameters,  it is possible to
work out a result for this integral exact in $\eps$, which is 
\begin{align}
B_2 &= 
\frac{\Gam(1-2\e)\, \w}{\e^2\,\Gam(1-\e)^2}\,
\Big( {}_2F_1(1,1+\eps;1-\eps;\w) - {}_2F_1(1,1;1-\eps;\w) \Big)
\nn&\quad
+ \frac{\w^{1-\eps}}{\eps^2}\, {}_2F_1(1,1-\eps;1-2\eps;\w) ~.
\label{eq:B2_fwd}
\end{align}
The computation of the physical cut $\Cut_1B_2$ (see \cref{fig:one-loop-crossed-box-cutting-equation}) is essentially the same as the cut diagram of the normal box $B_1$, since it again has a two-particle phase space non-vanishing for $s>Q^2$. 
It reads
\begin{align}
\Cut_1 B_2&=
\loopF \,Q^{2\e}\,s^2\,(-2 \pi i)^2
\int\d^{4-2\eps}k\frac{\delta^+(k^2)\delta^+((p+\pbar-k)^2-Q^2)}{(k-p)^2\,(k-\pbar)^2} \nn
&=
- 4 i \pi \,
z^{\eps}\,(1-z)^{-1-2\eps}
\frac{\theta(z)\,\theta(1-z)}{\eps\,\Gam(1-\eps)\,\Gam(1+\eps)}~.
\label{eq:B2_cut1}
\end{align}
In order to clarify the structure of the cutting equation,
we explicitly compute also
the unphysical $u$-channel cut diagrams (the second diagram on the right-hand side of \cref{fig:one-loop-crossed-box-cutting-equation}).
This is defined as
\begin{align}
\Cut_2 B_2
&= \loopF Q^{2\e}\,s^2\, (-2 \pi i)^2 
\int\d^{4-2\eps}k
\frac{\delta^+(k^2)\delta^+((p-\pbar-k)^2)}{(k-p)^2\,[(k+\pbar)^2-Q^2]} ~.
\end{align}
In contrast with the $s$-channel cut, this is non-vanishing when $s<0$, so we  perform the computation in the following frame:
\begin{align}
p = \frac{\sqrt{-s}}{2} (1,1,0,0) ~,~~
\pbar = \frac{\sqrt{-s}}{2} (-1,1,0,0) ~,~~
k = k^0 (1,\cos\theta,\sin\theta,0) ~.
\end{align}
Putting the right-most vertical propagator on-shell fixes $k^0 = \frac{\sqrt{-s}}{2}$. 
The calculation in this frame yields
\begin{align}
\Cut_2 B_2
&= 2 \pi i \,e^{- i \pi \eps}\,
\frac{\Gam(-\eps)}{\Gam(1-\eps)^2\,\Gam(1+\eps)}\,
z^{-1+\eps}\,
{}_2F_1(1,1-\eps;1-2\eps;1/z)\,
\theta(-z)~.
\label{eq:B2_cut2}
\end{align}
Combining \cref{eq:B2_fwd,eq:B2_cut1,eq:B2_cut2}, it is straightforward to verify that
\begin{align}
\underset{\w}{\Disc}\, B_2= \Cut_1 B_2 + \Cut_2 B_2 ~,
\end{align}
as shown in \cref{fig:one-loop-crossed-box-cutting-equation},
proving that in $z$-space both cuts are needed to reproduce the discontinuity. 
Hence, in order to work out the physical cut contribution, one would need to subtract the unphysical cut from the discontinuity of the forward diagram. 
As we shall see, this can be bypassed in Mellin space by using the shifting procedure
that we introduced in \cref{sec:one-loop-triangle} for the triangle diagram.

We start writing \cref{eq:B2_fwd}  as a series representation for the hypergeometric functions and moving the 
overall ${\w}$ inside the sums. 
This gives
\begin{align}
B_2 &=
\frac{\Gam(1-2\eps)}{\eps^2\,\Gam(1-\eps)} 
\Bigg[
\sum_{n=1}^{\infty} 
\frac{ - \Gam(n)}{\Gam(n-\eps)} \w^n 
+ \sum_{n=1}^{\infty} 
\frac{1}{\Gam(1+\eps)} 
\frac{\Gam(n+\eps)}{\Gam(n-\eps)} \w^n 
+ \sum_{n=1}^{\infty} 
\frac{\Gam(n-\eps)}{\Gam(n-2\eps)} \w^{n-\eps}
\Bigg] \,.
\label{eq:B2_three_sums}
\end{align}
We note that the last term contains a non-integer power of $\w$, which prevents us from constructing a formula for the series coefficients of the forward amplitude. 
Therefore, we apply to this term the same trick that
we used for the triangle diagram.
We change $n \to n+\eps$ in the summand 
but not in the range of the sum (i.e. we sum from $n=1$).
This gives for the last term
\begin{align}
\sum_{n=1}^{\infty} 
\frac{\Gam(n)}{\Gam(n-\eps)} \w^{n}~.
\end{align}
We have now defined a \emph{new} function $\widetilde B_2$ with no branch cut starting at the origin, from which we can extract the series coefficients. 
Applying this prescription to \cref{eq:B2_three_sums}, we see that last term cancels against the first sum and we are left with the second term. 
In conclusion, writing $\widetilde B_2 = \sum_{n} \widetilde c_n \w^n$, we have
\begin{align}
\widetilde c_n &=
\frac{\Gam(1-2\eps)}{\eps^2\,\Gam(1-\eps)\,\Gam(1+\eps)} 
\frac{\Gam(n+\eps)}{\Gam(n-\eps)} ~.
\label{eq:B2_fwd_mom}
\end{align}
Now we move to the cut diagrams.
As for the normal box $B_1$, they can be computed after writing them as function of $z=Q^2/s$, and then performing the standard Mellin transform. 
Using the results in \cref{eq:B2_cut1,eq:B2_cut2} this yields
\begin{align}
\Mellin{\Cut_1 B_2}
&= 2 \pi i \,\,
\frac{\Gam(1-2\eps)}{\eps^2\,\Gam(1-\eps)\,\Gam(1+\eps)}
\frac{\Gam(n+\eps)}{\Gam(n-\eps)}~, \nn
\Mellin{ \Cut_2 B_2 }
&= 0 ~.
\label{eq:B2_cut_mom}
\end{align}
The latter moments are zero due to the step function $\theta(-z)$.
Comparing the series coefficients in \cref{eq:B2_fwd_mom} to the moments of the physical cut in the first line of \cref{eq:B2_cut_mom}, we see that
\begin{align}
\widetilde c_n = \frac{1}{2\pi i}  \Mellin{\Cut_1 B_2}~,
\end{align}
which 
verifies \cref{eq:cutting_Mellin}.
We conclude that the Mellin moments of the physical cut are indeed reproduced by the series coefficients of the modified forward amplitude $\widetilde B_2$ and 
that the unphysical cut has been removed by the shifting procedure.

\subsection{Shifting procedure}
\label{sec:shifting-procedure}
For the triangle and the crossed-box diagrams,
we introduced a prescription to deal with a forward amplitude $f(\w)$ that cannot be expanded around $\w=0$. 
We also saw that for the crossed box diagram this is due the presence of a non-vanishing massless $u$-channel cut, which encodes the part of the forward amplitude having a branch cut along the \emph{negative} real axis in the $\w$-plane. For the triangle diagram, instead,
this is due to a non-vanishing massless $s$-channel cut,
with branch cut in the \emph{positive} real axis in the $\w$-plane. 

In order to clarify the shifting procedure let us review its general features.
Assume that (a non-analytic piece of) the forward amplitude $f(\w)$ can be written as
\begin{align}
f(\w) = \w^{k \e} g(w)~,
\label{eq:shifting_procedure_forward}
\end{align}
where $k$ is some integer and $g(\w)$ is analytic in $\w=0$.
The following discussion will trivially generalise to the case with more terms, which for instance might have different values of $k$ or different signs like $(\pm \w)^{k \eps}$.
The functions $f(\w)$ and $g(\w)$ may also depend on $\eps$, which is left implicit for brevity.
Writing $g(\w)$ as an expansion around $\w=0$, we have
\begin{align}
f(\w)=\sum_{n=n_0}^{\infty} c_n\,\w^{n+k\eps}~.
\label{eq:shifting_procedure_forward_series}
\end{align}
The \emph{shifting procedure} is defined by replacing $n$ with $n-k\eps$ in the summand, but not in the lower bound of the sum.%
\footnote{
An alternative definition of the shifting procedure is the following. First rewrite the forward amplitude as
$f(\w) = \sum_{n=n_0+k\eps}^{\infty} c_{n-k\eps} \,\w^{n} $, where sums starting at non-integer lower bound $\alpha \in \mathbb{C}$ are to be interpreted as $\sum_{n=\alpha}^{\infty} s_n = s_{\alpha} + s_{\alpha+1} + \dotsm$. The shifting procedure may then alternatively be defined as setting $\eps = 0$ in the lower bound of the sum, 
so that $f(\w) \to \widetilde f(\w) = \sum_{n=n_0}^{\infty} c_{n-k\eps} \,\w^{n}$.
}
This produces a new function
\begin{align}
\widetilde f(\w) = \sum_{n=n_0}^{\infty} c_{n-k\eps}\,\w^{n}~.
\label{eq:shifting_procedure_forward_tilde}
\end{align}
This function $\widetilde f(\w)$ is a modified version of the forward diagram that is precisely what is needed for our purpose, if the following two criteria are met:
\begin{enumerate}[label=(\roman*)]
\item The unphysical cut must be absent in $\underset{\w}{\Disc}\, \widetilde f(\w)$;
\item The discontinuity around the physical cut must be unaffected.
\end{enumerate}
We discuss the validity of these two conditions in turn.

The first criterium goes back to the assumption made in \cref{eq:shifting_procedure_forward}, namely that the non-analyticity of $f(\w)$ around $\w=0$ is captured by an overall factor $\w^{k\eps}$.
This can be argued with dimensional analysis, when looking at the physical complex $s$-plane (remembering that $\w = s/Q^2$). 
Branch cuts starting at $s=0$ are described by the single dimension-full quantity $s$, irrespective of the value of $Q^2$.
Since Feynman diagrams have a fixed integer mass dimension, the only way in which $s$ can occur is as an overall power of $s$ and not as the argument of some other (elementary) function. 
Fractional powers are excluded by dimensional analysis.
The only deviation from integer $s$ powers allowed is due to the $d$-dimensional integration measure.
Feynman integrals yield results proportional to $s^{k\eps}$ which, combined with the dimensional regularisation mass scale set to $Q^2$, produces overall factors $\w^{k\eps}$.
We thus conclude that any unphysical cut starting at $\w = s = 0$ is captured by functions of the form in \cref{eq:shifting_procedure_forward_series}.
The modified forward amplitude in \cref{eq:shifting_procedure_forward_tilde} is analytic at $\w = 0$ by construction and therefore the unphysical cut is indeed completely removed from $\Disc f(\w)$ by the shifting procedure.

The second criterium ensures that altering the forward amplitude, does not destroy the connection between the discontinuity around the physical branch of the forward amplitude and the sum of physical cut diagrams.
This issue can be clarified through some toy examples. 
Let us first consider a simple case where the ``forward amplitude'' is given by
\begin{align}
f_1(\w) 
= \frac{\w^{-\eps}}{1-\w}
= \sum_{n=0}^{\infty} \w^{n-\eps}~.
\end{align}
Applying the shifting procedure, we arrive at the new function
\begin{align}
\widetilde{f}_1(\w) 
= \sum_{n=0}^{\infty} \w^{n} 
= \frac{1}{1-\w}~.
\end{align}
While the original function has its branch cut along the negative real axis removed, both functions have the same pole structure in the region $\w \geq 1$, namely 
\begin{align}
\underset{\w \geq 1}{\Disc}\, f_1(\w) 
= 2 \pi i \, \delta(1-\w) 
= \underset{\w \geq 1}{\Disc}\, \widetilde f_1(\w)  ~.
\end{align}
Therefore in this example the discontinuity in the physical region is unaltered by the shifting procedure. 
Another example is given by a function with a branch cut, rather than a simple pole, in the physical region. 
This mimics more closely the cases we encountered at one loop.
Consider
\begin{align}
f_2(\w) 
= - \log(1-\w) \, \w^{-\eps} 
= \sum_{n=1}^{\infty} \frac{1}{n} \,\w^{n-\eps} ~.
\end{align}
The shifting procedure produces
\begin{align}
\widetilde f_2(\w)
= \sum_{n=1}^{\infty} \frac{1}{n+\eps} \, \w^{n} ~.
\end{align}
Again, these functions have the same discontinuity in the physical region. 
This is most easily verified upon writing both functions as an expansion in $\eps$,
\begin{align}
f_2(\w) &= -\log(1-\w) \sum_{k=0}^{\infty} (-\eps)^k \, \frac{\log^k(\w)}{k!} ~,\nn
\widetilde f_2(\w) &= \sum_{k=0}^{\infty} (-\eps)^k \, \Li_{k+1}(\w) ~.
\end{align}
Using the identities
\begin{align}
&\underset{\w \geq 1}{\Disc} \,\big[\! \log(1-\w) \log^{k}(w) \big] = - 2 \pi i \, \log^{k}(\w) ~, \nn
&\underset{\w \geq 1}{\Disc} \,\big[ \Li_{k+1}(\w) \big] = 2 \pi i \, \frac{\log^{k}(\w)}{k!} ~,
\end{align}
one readily confirms that the discontinuity of $f_2(\w)$ in the physical region is equal to that of $\widetilde f_2(\w)$ and is thus unaltered by the shifting procedure.

\subsection{Direct extraction of series coefficients from IBP's}
\label{sec:extraction-of-series-coefs}

The shifting procedure from the previous subsection requires the result of an integral to be given in terms of $\omega^{-\eps}$ in \emph{unexpanded} form. 
Indeed, if the integral were expanded in $\eps$, then the logarithmic divergence at $\omega=0$ could no longer be removed by shifting $n$.
In the one-loop examples of the previous section, there is no problem
since the forward amplitude diagrams are exact in $\eps$. 
Moreover, for each of the one-loop diagrams a simple series
representation is known, which allows us to extract their series coefficients exact in $\eps$ as well.

At higher loops it is not realistic to expect exact results in $\eps$ for all the forward diagrams.
However, it is in fact sufficient that the divergent part of a forward
diagram $f(\omega,\eps)$ around $\omega=0$ is written as
$\omega^{-\eps} \, g(\omega,\eps)$, where $g(\omega,\eps)$ is regular
at $\w=0$,
and may also be given in expanded form $g(\omega,\eps) = \sum_{k \geq k_{0}} \eps^k \, g_k(\omega)$.
Such a hybrid expression can be obtained by making a series ansatz for the forward diagram.
Specifically, for one-loop diagrams one writes 
\begin{align}
f(\omega,\eps) = s^{\text{dim}[f]} \left( \sum_{n} c_n(\eps) \, \omega^n + \omega^{-\eps} \sum_n d_n(\eps) \, \omega^n \right) ~,
\label{eq:one_loop_ansatz_for_forward}
\end{align}
where $\text{dim}[f]$ denotes the integer mass dimension of the forward amplitude $f$.
This structure for a forward diagram is not surprising, since
in general the function is non-analytic at the origin $\omega=0$.
Without loss of generality, one can decompose such a function into a sum of analytic and non-analytic pieces. 
As discussed in the previous subsection, the non-analyticity can always be captured by a factor $\omega^{-\eps}$ multiplied by another function, which is regular at the origin and thus admits a series representation. 
The coefficients $c_n(\eps)$ and $d_n(\eps)$ may be given exact in $\eps$ or as an expansion in $\eps$; in either case the shifting procedure works.

A further benefit of making the series ansatz in \cref{eq:one_loop_ansatz_for_forward} is that the series coefficients can be extracted more directly by deriving equations for them and subsequently solving the equations.
One way to proceed along these lines is to generate a differential equation for $f(\omega,\eps)$ from integration-by-parts (IBP) identities.
Inserting the series ansatz into such differential equation yields in turn a \emph{difference} equation for the series coefficients, which takes the form
\begin{align}
A_{0,n} \, c_{n}(\eps) + A_{1,n} \, c_{n+1}(\eps) + \dotsm + A_{r,n} \, c_{n+r}(\eps) = F_{n} ~,
\label{eq:generic_difference_equation}
\end{align}
where the $A_{i,n}$ and $F_n$ are rational functions of $\eps$, whose form depends on the particular differential equation for $f(\w,\eps)$ under consideration.
If $r=1$, then \cref{eq:generic_difference_equation} is simply a recursion, in which case the series coefficients can be found exact in $\eps$.
For diagrams with multiple loops the order of the difference equation becomes typically quite high (we find up to $r=8$ for two-loop diagrams).
In that case it will be advantageous to expand the difference equation \cref{eq:generic_difference_equation} in $\eps$ and solve for the coefficients $c_{n}(\eps) = \sum_{k \geq k_{0}} \eps^k \, c_{k,n}$, order-by-order in $\eps$.
The task of solving the resulting difference equations for the set
$\{c_{k,n}\}$ may be even further simplified by following the approach
in \citeR{Vermaseren:2000we}, which exploits the expectation that
these coefficients are given in terms of harmonic numbers.

For the computation of the two-loop diagram in this paper we indeed adopt the approach in \citeR{Vermaseren:2000we} and seek solutions to \cref{eq:generic_difference_equation} of the form
\begin{align}
c_{n}(\eps) = \sum_{k,\vec\ell,m} A_{k,\ell,m} \, \eps^k \, S_{\vec\ell\,}(n-m) ~.
\label{eq:ansatz_series_coefficients}
\end{align}
for reasonable choices of $k,\vec\ell,m$.
The functions $S_{\vec\ell\,}(n)$ are harmonic sums, whose properties are well-known \cite{Vermaseren:1998uu}.
The unknown coefficients $A_{k,\vec\ell,m}$ contain both rational and
transcendental numbers, and may be determined as follows.
The simplest approach is to evaluate the difference equation for $c_{k,n}$ at suitably many values of $n$, so as to obtain a system of equations which may be solved for the unknown $A_{k,\vec\ell,m}$. 
In a more sophisticated method \cite{Vermaseren:2000we} each term in the difference equation for $c_{k,n}$ is projected onto a basis of synchronised harmonic numbers, after which the coefficients of each harmonic number is equated to zero. 
This also yields a system of equations for the unknown $A_{k,\vec\ell,m}$.
We have implemented both techniques and successfully applied them to the two-loop examples in \cref{sec:two-loop-examples}.
\newline

Before closing this section, let us present an example of the methods in this subsection for obtaining the series coefficients from IBP's.
To this end, consider the one-loop box $B_1$ from \cref{sec:one-loop-normal-box}.
After shifting the loop momentum $k \rightarrow q = k + \pbar$ in \cref{eq:B1_fwd}, the scalar box integral becomes a special case $B(2,1,1)$ of the topology
\begin{align}
B(a,b,c) &= \int\frac{\d^{4-2\eps}q}{(q^2)^a \, ((q+p)^2)^b \, ((q-\pbar)^2-Q^2)^c} ~.
\label{eq:B(abc)_topology}
\end{align}
The derivative of $B(a,b,c)$ with respect to $Q^2$ produces another integral in the topology:
\begin{align}
\frac{\partial}{\partial Q^2} B(a,b,c) &= c \, B(a,b,c+1) ~~\text{for}~c \geq 0~.
\label{eq:derivative_B(abc)}
\end{align}
Performing IBP reduction on the right-hand side of \cref{eq:derivative_B(abc)} thus yields a differential equation for $B(a,b,c)$ in terms of (typically) simper integrals. 
In the case of the one-loop box, the differential equation for $B(2,1,1)$ reads%
\footnote{
This equation directly follows after inserting $\frac{\d}{\d q^\mu} (q+p)^\mu$ under the integral sign in \cref{eq:B(abc)_topology} with $(a,b,c)=(2,1,1)$ and using \cref{eq:derivative_B(abc)}.
}
\begin{align}
\bigg[(-1-2 \eps)+(s-Q^2) \frac{\partial}{\partial Q^2} \bigg] B(2,1,1)
&= 2 B(3,0,1) + \frac{\partial}{\partial Q^2} B(2,0,1) ~.
\end{align}
The inhomogeneous terms on the right-hand side are simpler integrals because they have fewer propagators. 
Inserting the known bubbles $B(a,0,1)$ on the right-hand side and the definition of $B(2,1,1)$ in terms of the forward box $B_1$ on the left-hand side, leads to a differential equation for $B_1$:
\begin{align}
\bigg[ (-1 -\eps -\eps\,\omega)+(s-Q^2) \frac{\partial}{\partial Q^2} \bigg] B_1
= \frac{\Gam(1-2\eps) }{ \eps \, \Gam(1-\eps)^2} \frac{1}{(Q^2)^{2}} ~.
\end{align}
One way to solve this differential equation is by inserting an ansatz for $B_1$, like in \cref{eq:one_loop_ansatz_for_forward}, 
in terms of unknown coefficients $c_n(\eps)$ and $d_n(\eps)$.
Upon doing so, one finds that $d_n(\eps) = 0$, while the other coefficients $c_n(\eps)$ satisfy the equation 
\begin{align}
c_{2}(\eps) \, \omega^2 (1-\eps) +
\sum_{n=3}^{\infty} \bigg[ (n-1-\eps) \, c_{n}(\eps) - (n-1+\eps) \, c_{n-1}(\eps) \bigg] \omega^{n}
= \frac{\Gam(1-2\eps)}{\eps\,\Gam(1-\eps)^2} \, \omega^2 ~.
\end{align}
Equating terms with equal powers of $\omega$ on both sides leads to a recurrence relation in $n$,
\begin{align}
c_{2}(\eps) = \frac{\Gam(1-2\eps)}{\eps \, \Gam(1-\eps) \, \Gam(2-\eps)} ~,\quad
c_{n}(\eps) = \frac{n-1+\eps}{n-1-\eps} \, c_{n-1}(\eps)  \quad \text{for}~ n > 2 ~.
\end{align}
This recursion is solved by
\begin{align}
c_{n}(\eps) &= \frac{1}{\eps\,(1+\eps)}\frac{\Gam(1-2\eps)}{\Gam(1-\eps)\Gam(1+\eps)}\frac{\Gam(n+\eps)}{\Gam(n-\eps)} ~.
\end{align}
This result for the series coefficients of $B_1$ fully agrees with \cref{eq:B_1_series}.

In this simple example it has been possible to solve the recursion exact in $\eps$.
For higher loop diagrams this is typically not expected to be possible to do
exactly but rather order-by-order in $\eps$. 
In the next section we extend our investigations to two-loop diagrams, where it is shown how to solve a higher-order difference equation by making an ansatz for the series coefficients in terms of harmonic numbers.

\section{Drell-Yan at two loop}
\label{sec:DY-two-loop}

The previous section discussed for Drell-Yan production at the one-loop level, how Mellin moments of cut diagrams can be computed as series coefficients of forward diagrams.
A feature in our approach is that unphysical cuts in forward amplitude diagrams can be removed by a shifting procedure.
At higher loops this shifting procedure is no longer sufficient. 
Indeed, in this section we extend our investigations to two loops, for which we develop an additional prescription to subtract unphysical cuts.
Two-loop diagrams serve furthermore as non-trivial applications of our method for direct extraction of Mellin moments from integration-by-parts identities, as discussed in \cref{sec:extraction-of-series-coefs}.
In the next subsection we start by listing all possible types of unphysical cuts, placing particular emphasis on the new type of unphysical cut appearing at two loops.
We then describe our methods to remove them, working out two examples in detail.

\subsection{Unphysical cuts of two-loop diagrams}
\label{sec:unphysical-cuts-two-loops}

\begin{figure}[t]
\begin{center}
\raisebox{3mm}{
\includegraphics{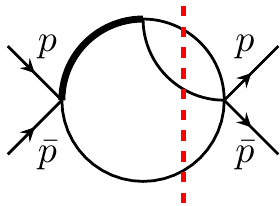}
}
\hspace{10mm}
\includegraphics{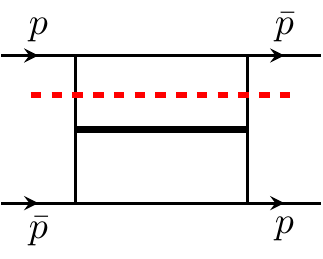}
\hspace{10mm}
\includegraphics{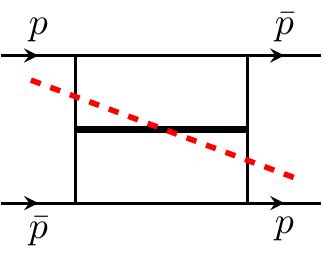}
\\ 
\vspace{-3mm}
\flushleft
\hspace{31mm}(a)
\hspace{38mm}(b)
\hspace{40mm}(c)
\caption{Types of unphysical cuts appearing at two loops. Example are: 
(a) massless $s$-channel cut;
(b) massless $u$-channel cut;
(c) massive $u$-channel cut.
}
\label{fig:two-loop-unphysical-cuts}
\end{center}
\end{figure}

Let us analyse the possible unphysical cuts of two-loop forward diagrams.
At one-loop level there are two types: unphysical $s$-channel cuts and massless $u$-channel cuts (which do not cut the massive photon).
At two loops, there is the possibility for a new type: massive $u$-channel cuts (which \emph{do} cut the massive photon). 
All types of unphysical cuts are illustrated in \cref{fig:two-loop-unphysical-cuts}.
We briefly discuss the differences between these types of unphysical cuts.

\paragraph{Massless $s$-channel cuts}
A cut of this type appears already in the case of the one-loop triangle in \cref{fig:one-loop-triangle-cutting-equation}.
A two-loop example is given in \cref{fig:two-loop-unphysical-cuts}a, which features a three-particle massless phase-space integral. 
In general, diagrams in this category are always massless phase-space integrals, because the massive line is not cut by definition.
Such massless phase-space integrals always come with a step function $\theta(s)$, which indicates that it arises from the discontinuity around a logarithmic branch cut starting at the origin $s=0$ (or $\w=0$). 
This situation is reminiscent of the unphysical branch cut corresponding to massless $u$-channel cut diagrams, which can be removed by the shifting procedure from \cref{sec:shifting-procedure}. 
Indeed, we find that the shifting procedure is sufficient to deal with these unphysical $s$-channel cuts, which is supported by the example to be treated in \cref{sec:two-loop-real-virtual}.

\paragraph{Massless $u$-channel cuts}
Cuts in the $u$-channel are for our case unphysical by definition,
as they do not occur in the cut-diagrammatic expansion of the Drell-Yan cross section.
The simplest class of unphysical $u$-channel cuts are the ones where only massless lines are cut.
Examples of such cuts are depicted in \cref{fig:one-loop-crossed-box-cutting-equation} and \cref{fig:two-loop-unphysical-cuts}b, in the case of one- and two-loop crossed-box diagrams, respectively.
These cuts correspond to branch cuts of forward diagrams with the branch point at the origin, 
so they can be removed by the shifting procedure from \cref{sec:shifting-procedure}.
In \cref{sec:two-loop-crossed-box} a two-loop example is discussed where this procedure is applied.

\paragraph{Massive $u$-channel cuts}
This is a new type of $u$-channel cut which first appears at the two-loop
level. The presence of the massive line has the effect of shifting the branch point to $\w=-1$, 
as compared to situation of the massless $u$-channel cuts.
In this case the shifting procedure cannot be applied, 
so new method must be introduced to remove this type of unphysical cut.
In the next section we focus on this problem and propose a solution in the form of an extra prescription.
A non-trivial test of that procedure is then presented in the context of the two-loop crossed box in \cref{sec:two-loop-crossed-box}. 
\newline

\begin{figure}[t]
\begin{center}
\includegraphics{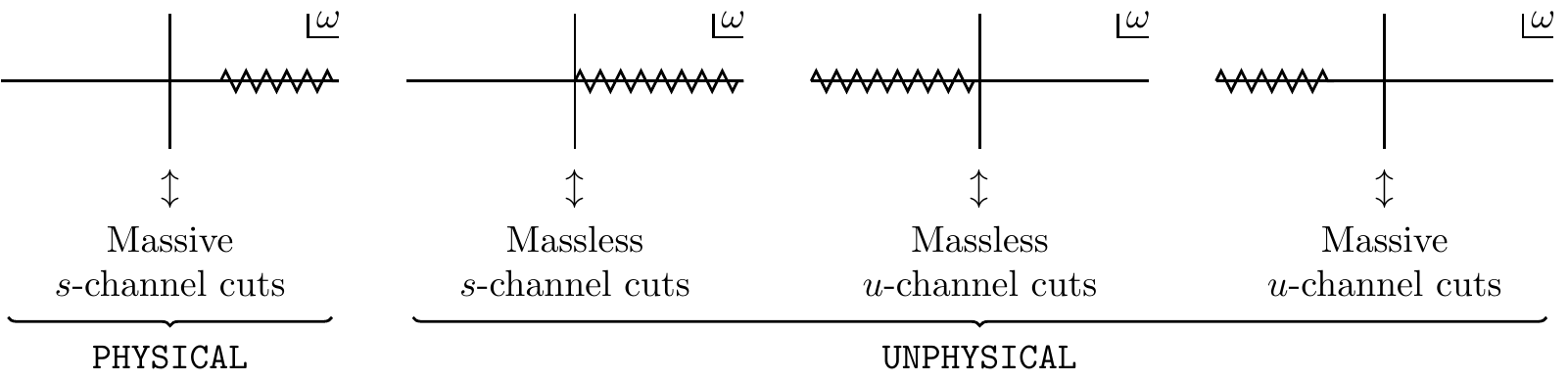}
\caption{Branch cut structures that appear in two-loop forward diagrams, listed together with cuts that describe the corresponding branch cut discontinuities according to the cutting equation. Only the first type of cut is physical, i.e. contributes to the Drell-Yan cross section.}
\label{fig:analytic-structure-unphysical-cuts}
\end{center}
\end{figure}

The various types of unphysical cuts correspond to branch cut discontinuities of forward amplitude diagrams, where the branch cut does \emph{not} extend from $\w=1$ to $\w=\infty$. 
The connection between the above-mentioned cut diagrams and branch cut discontinuities is summarized in \cref{fig:analytic-structure-unphysical-cuts}.

\subsection{Extracting series coefficients for two-loop forward
  amplitude diagrams} 
\label{sec:approach-two-loop-amplitudes}

The analysis of the types of unphysical cuts of two-loop diagrams in the previous subsection calls for an extension of our method for extracting the series coefficients of forward amplitudes, as described in \cref{sec:DY-one-loop} for one-loop diagrams. 
In particular, the massive $u$-channel cut requires a new prescription besides the shifting procedure in \cref{sec:shifting-procedure}. 
The shifting procedure itself works also at higher loops, but the series ansatz for a forward diagram in \cref{eq:one_loop_ansatz_for_forward} is particular to the one-loop case and needs to be generalised.
Furthermore, when dealing with higher-loop diagrams, the discussion in \cref{sec:extraction-of-series-coefs} on how to obtain the series coefficients from IBP's should be combined with the notion of master
integrals. 
We start by discussing the latter.  

In many calculations of scattering amplitudes at higher orders in the QCD coupling constant, the use of master integrals has proven to be extremely useful.
There can be many diagrams contributing to a cross-section, which produce even more Feynman integrals upon working out tensor reduction.
Typically, all those integrals can be written as special cases of a handful of \emph{topologies}: integrals with as many linearly independent propagators as Lorentz invariants formed out of at least one loop momentum, raised to arbitrary powers. 
It has been shown that all integrals in a topology can be written in terms of a finite set of \emph{master integrals} \cite{Lee:2013hzt}.
The computation of Feynman integrals for cross-sections thus boils down to computing master integrals.
For this reason we focus in the remainder of this section on applying our method to master integrals.

Let $\mathbf{M}$ denote a vector of $n$ such master integrals which depend on $\w$ and $\eps$.
Assume that the first $k$ master integrals can be computed by applying known analytical formulae for one-loop bubbles successively.
We indicate these with a superscript $B$.
Then the vector of master integrals is written as
$\mathbf{M} = (M^{B}_{1},\dotsc,M^{B}_{k},M_{k+1},\dotsc,M_{n})$.
Given the fact that the bubble-type integrals $M^{B}_{i}$ are known exactly in $\eps$, 
they can serve as inhomogeneous terms for the differential equations for the remaining unknown $M_i$. 
This works as follows.
Gathering the unknown integrals in the vector $\mathbf{\overline{M}} = (M_{k+1},\dotsc,M_{n})$, taking its derivative with respect to $\w$ and reducing the result to master integrals yields a system of first-order coupled differential equations $\frac{d}{d\w} \mathbf{\overline{M}} = \mathbf{A} \cdot \mathbf{M}$.
Notice here that the right-hand side generally depends on \emph{all} master integrals.
This situation can be avoided by decoupling the differential equations, at the expense of raising the order of the differential equations \cite{Gehrmann:1999as}.
As a result, the differential equation for a given $M_i$ will then be
of order $r$, which takes some value $1 \leq r \leq k$ depending on
the particular system, and has the form
\begin{align}
\left(\sum_{m=0}^{r} a_m \frac{d^m}{d\w^m} \right) M_i 
= \left(\sum_{j=1}^{k} \sum_{m=0}^{r-1} b_{j,m} \frac{d^m}{d\w^m}\right) M^B_j ~.
\label{eq:differential_equation_Mi}
\end{align}
Here, the free index $i$ is bound by $k+1 \leq i \leq n$ and the coefficients $a_m$ and $b_{j,m}$ are rational functions of $\w$ and $\eps$.
We emphasise that the right-hand side is known exactly in $\eps$, since it consists of the known bubble-type integrals and derivatives thereof.
The number of unknown integrals $k$ can be large, in practice.
This means that the order of the differential equation $r \in [1,k]$ could be equally large, making it challenging to solve.
Moreover, the rational functions $a_m$ and $b_{j,m}$ also grow in size as $r$ increases.
Such situations may be ameliorated by decoupling differential equations to subsets of master integrals in $\mathbf{\overline{M}}$.
After each iteration the solutions that can be found exact in $\eps$ may be used as inhomogeneous terms as well, thus lowering the order of the differential equations for the next integrals to be calculated.

The next task is to solve the differential equations in \cref{eq:differential_equation_Mi} for $i=k+1,\dots ,n$.
The approach we shall take in this paper is that of inserting an ansatz for the $M_i$ in terms of series expansions in $\w$.
The one-loop ansatz in \cref{eq:one_loop_ansatz_for_forward} already displays the key feature, namely of decomposing the function into analytic and non-analytic pieces.
The non-analyticity at $\w=0$ is captured by powers $\w^{-\eps}$.
In the case of two-loop diagrams the ansatz will take the following form
\begin{align}
M_i = 
\sum_n c^{(i)}_n \, \omega^n 
+ \sum_n d^{(i)}_n \, \omega^{n-\eps}
+ \sum_n e^{(i)}_n \, \omega^{n-2\eps} ~,
\label{eq:two_loop_ansatz_for_forward}
\end{align}
where the series coefficients $c^{(i)}_n,d^{(i)}_n, e^{(i)}_n$ depend on $\eps$ but not on $\w$.
Substituting this expression for $M_i$ into the differential equation \cref{eq:differential_equation_Mi} and equating equal powers of $\w$ produces difference equations for the series coefficients.
The difference equations have the general form of \cref{eq:generic_difference_equation}.
Non-zero coefficients of $\w^k$ on the right-hand side of the differential equation supply boundary conditions to the difference equations. 
This means that additional computations to ascertain such boundary conditions, e.g. using expansion-by-regions, are (typically) not necessary.
In simple cases, when the order of the difference equation is relatively low, the series coefficients might be solved exactly in $\eps$, involving ratios of Gamma-functions.
Otherwise, the series coefficients can always be solved 
order-by-order in $\eps$ in terms of harmonic sums, using an ansatz of
the form given in \cref{eq:ansatz_series_coefficients}.
Once one has the series coefficients $c^{(i)}_n,d^{(i)}_n, e^{(i)}_n$ in hand, we have essentially computed the forward diagram.

Then we are in the position to start modifying the forward diagram in such a way that its discontinuity no longer has any contribution from unphysical cuts.
The first step in this process is the shifting procedure.
Conceptually, this procedure prescribes exactly what was discussed for one-loop diagrams: terms in the series are replaced according to $c_n\,\w^{n-k\eps} \rightarrow c_{n+k\eps}\,\w^n$.
After shifting, a two-loop forward master integral thus becomes
\begin{align}
M_i \rightarrow \widetilde{M}_i =
\sum_n \left( c^{(i)}_n + d^{(i)}_{n+\eps} + e^{(i)}_{n+2\eps} \right) \omega^{n} 
\equiv \sum_n \widetilde{c}^{\,(i)}_n \, \omega^{n} ~.
\label{eq:two_loop_ansatz_shifted}
\end{align}
As discussed in \cref{sec:shifting-procedure}, this procedure removes the unphysical branch cut discontinuity that arise from massless $s$-channel cuts and/or massless $u$-channel cuts.
Indeed, the function $\widetilde{M}_i$ is expanded as a series around $\w=0$.
At a technical level, this shifting procedure can be a bit more involved than the one-loop case.
Namely, if the coefficients $c^{(i)}_n,d^{(i)}_n, e^{(i)}_n$ were expressed order-by-order in $\eps$ in terms of harmonic numbers, then the series in \cref{eq:two_loop_ansatz_shifted} features harmonic numbers evaluated at non-integer values: $S_\ell(n+k\eps)$.
These functions must be expanded in $\eps$ to match the form of the rest of the expression, which boils down to taking derivatives of harmonic numbers with respect to their argument.
To this end one uses the known analytic continuation of harmonic numbers from the integers to the real line.
In practice, we make use of the package \texttt{HarmonicSums} \cite{%
Ablinger:2010kw,%
Ablinger:2013hcp,%
Ablinger:2013cf,%
Ablinger:2011te,%
Blumlein:2009ta,%
Remiddi:1999ew,%
Vermaseren:1998uu%
} to expand the harmonic numbers evaluated at non-integer values.
\newline

So far we have discussed the shifting procedure and the use of IBP's to extract series coefficients in the context of two-loop diagrams.
This is sufficient to deal with all types of unphysical cut of two-loop diagrams, \emph{except} for massive $u$-channel cuts.
The latter type of cut diagrams correspond to a branch cut of the forward along $(-\infty,-1]$ in the complex $\w$-plane.
Since the branch point is not at the origin $\w=0$, it is not removed by the shifting procedure.
In order to remove discontinuities around such branch cuts, we extend
our method further.

\begin{figure}[t]
\begin{center}
\includegraphics{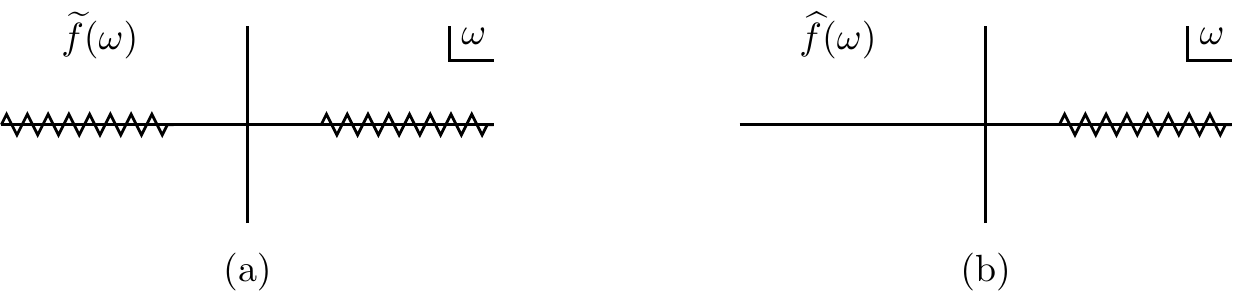}
\caption{
The analytic structure of the example functions $\widetilde{f}(\w)$ and $\widehat{f}(\w)$, given in \cref{eq:example_f} and \cref{eq:example_f_tilde_result}, respectively. The function $\widetilde{f}(\w)$ represents a forward diagram, whose discontinuity contains contributions from both physical and un-physical cut diagrams. The second function, $\widehat{f}(\w)$, is a modified version of the forward, such that only the physical branch cut is present. 
}
\label{fig:analytic_structure_before_and_after_removing_wrong_branch_cut}
\end{center}
\end{figure}

We shall replace the forward diagram by a new function, whose branch cut along $\w \in (-\infty,-1]$ is removed while its branch cut discontinuity around the physical region $\w \in [1,\infty)$ remains unchanged.
Our technique for obtaining a function that satisfies these requirements is perhaps best explained with the help of an example.
Consider the following product of logarithms, denoted $\widetilde f(\w)$ in view of the absence of a branch point at $\w=0$,
\begin{align}
\widetilde{f}(\w) = \log(1+\w)\log(1-\w) ~.
\label{eq:example_f}
\end{align}
In the complex $\w$-plane this function $\widetilde{f}(\w)$ has two branch cuts, 
which are located along the disconnected intervals $(-\infty,-1]$ and $[1,\infty)$.
This situation is shown in \cref{fig:analytic_structure_before_and_after_removing_wrong_branch_cut}(a).
The discontinuity of $\widetilde{f}(\w)$ is simply the sum of the discontinuities around the individual branch cuts:
\begin{align}
\underset{\w}{\Disc}\, \widetilde{f}(\w) = 
2 \pi i \log(1-\w) \theta(-\w-1) 
- 2 \pi i \log(1+\w) \theta(\w-1) ~.
\label{eq:example_disc_f}
\end{align}
In \cref{eq:example_disc_f}, the first term on the right-hand side may be interpreted as a contribution coming from unphysical cut diagrams. 
Removing the unphysical cuts thus amounts to removing the branch cut along $(-\infty,-1]$ from the function $\widetilde{f}(\w)$, 
leaving a new function, $\widehat{f}(\w)$, such that 
\begin{align}
\underset{\w}{\Disc}\, \widehat{f}(\w) = 
- 2 \pi i \log(1+\w) \theta(\w-1) ~.
\label{eq:example_disc_f_tilde}
\end{align}
The corresponding analytic structure is displayed in \cref{fig:analytic_structure_before_and_after_removing_wrong_branch_cut}b.
The question is how to find $\widehat{f}(\w)$.

Note that in there is no unique answer to this question.
Indeed, any constant (or smooth function, for that matter) may be added without changing the discontinuity.
This ambiguity is lifted by imposing the constraint $\widehat{f}(0) = 0$, which reflects the physical property of scattering cross-sections that they vanish in the limit of zero centre-of-mass energy. 
This constraint, together with the analyticity of $\widehat{f}(\w)$ around the origin, 
allows us to write down a series representation
\begin{align}
\widehat{f}(\w) = \sum_{n=1}^{\infty} \widehat c_n \, \omega^n ~.
\label{eq:example_series_f_tilde}
\end{align}
The coefficients $\widehat c_n$ can be obtained from the Cauchy integral formula, taking a small contour enclosing the origin.
Inflating the contour such that it wraps around the branch cut, the contour integral becomes the integration of the discontinuity along the real line, analogous to the discussion in \cref{sec:DIS-optical-theorem}.
Subsequently changing variables to the reciprocal $z = 1/\w$ leads to the following Mellin-transform integral
\begin{align}
\widehat c_n = - \int_0^1 z^{n-1} \log(1+\tfrac{1}{z}) ~.
\label{eq:example_cn_Mellin_integral}
\end{align}
This standard integral transform may be performed (for more complicated cases one may use the \texttt{MT} package \cite{Hoeschele:2013gga}) and the result is
\begin{align}
\widehat c_n = - \frac{1}{n^2} + \frac{(-1)^n S_{-1}(n)}{n} - \frac{\log 2}{n} + \frac{(-1)^n \log 2}{n} ~.
\label{eq:example_cn}
\end{align}
In the analogy with perturbative computations, these coefficients correspond to the Mellin moments of the sum of cut diagrams obtained from the forward $\widetilde{f}(\w)$ by taking all possible physical cuts. 
Considering the aim of this paper, these moments therefore provide a satisfactory answer.

For completeness, we will also determine the full function $\widehat{f}(\w)$.
Obviously, in a small neighbourhood around the origin, $\widehat{f}(\w)$ is given by the series \cref{eq:example_series_f_tilde} with coefficients in \cref{eq:example_cn}.
Its analytic continuation to the complex $\w$-plane is given in terms of polylogarithms.
This continuation may actually be constructed by first rewriting the series coefficients as linear combination of harmonic sums with multiple indices \cite{Vermaseren:2000we}, which essentially projects $\widehat c_n$ onto a convenient basis of the function space:
\begin{align}
\widehat c_n = 
- \mathcal{S}_{2}(n) 
+ \mathcal{S}_{-1,-1}(n) 
+ \log 2 \,\, \mathcal{S}_{-1}(n) 
- \log 2 \,\, \mathcal{S}_{1}(n) ~,
\label{eq:example_cn_rewritten}
\end{align}
where $\mathcal{S}_{\ell}(n) = S_{\ell}(n) - S_{\ell}(n-1)$.
With this expression in hand, the sum in
\cref{eq:example_series_f_tilde} may be evaluated in closed form,
using the fact that the series coefficients of harmonic polylogarithms 
are harmonic numbers \cite{Remiddi:1999ew}, and one finds
\begin{align}
\widehat{f}(\w) 
&= - H_{-1,1}(\w) - \log 2 \,\, H_{-1}(\w) - \log 2 \,\, H_{1}(\w) \nn
&= - \Dilog\Big(\frac{1+\w}{2}\Big) + \log 2 \,\, \log(1-\w) - \frac{\log^2 2}{2} + \frac{\pi^2}{12} ~.
\label{eq:example_f_tilde_result}
\end{align}
One can check explicitly that this expression has the correct branch cut discontinuity, as required by \cref{eq:example_disc_f_tilde}. This completes the example. 

The same method for removing the unphysical branch cut can be applied to two-loop forward diagrams.
Apart from branch cuts, one then also deals with poles, typically at $\w=1$.
One simple way to implement the removal of the wrong branch cut is by deriving replacement rules for the individual harmonic numbers, which appear in the result of the shifting procedure \cref{eq:two_loop_ansatz_shifted}.
For a given harmonic number $S_{\vec{\ell}\,}(n)$, one first evaluates the corresponding sum $\sum_n S_{\vec{\ell}\,}(n)\,\w^n$ in closed form.
Based on similar analysis as in the previous example, one then constructs a function which has the unphysical branch cut removed and whose series coefficients define the replacement of $S_{\vec{\ell}\,}(n)$.
For example, in the case of $S_{1,-2}(n)$ we get
\begin{align}
\sum_{n=1}^{\infty} S_{1,-2}(n) \, \w^n 
= 
\frac{H_{-3}(\w) + H_{1,-2}(\w)}{\w-1}
~\to~
\frac{4 \, \zeta_2 \, H_{1}(\w) + \w \, \zeta_3}{8\,(\w-1)}
= \sum_{n=1}^{\infty} \Big(\!\! -\tfrac{1}{2} S_{1}(n)  - \tfrac{1}{8} \zeta_3 \Big) \, \w^n ~,
\end{align}
which is equivalent to the effective replacement rule $S_{1,-2}(n) \to -\tfrac{1}{2} S_{1}(n)  - \tfrac{1}{8} \zeta_3$.
Following these steps with all harmonic sums produces a `dictionary' of replacement rules, which may then be applied to any diagram.
We shall use such replacement rules in \cref{sec:two-loop-crossed-box}.

\subsection{Two-loop examples}
\label{sec:two-loop-examples}

This subsection provides examples that serve to illustrate two main lessons from our studies of two-loop diagrams, namely: how to remove unphysical cuts from a forward diagram, and how to compute the series coefficients of forward diagrams at higher loops from differential equations. 
The first example below will illustrate how to deal with massless $s$-channel cuts. 
We demonstrate that the shifting procedure not only removes massless $u$-channel cuts, but also removes
any unphysical $s$-channel cut. 
The second example shows the power of the method by applying it to a rather difficult forward amplitude diagram, the two-loop crossed box. 
The latter admits massive $u$-channel cuts, which can be treated along the lines of \cref{sec:approach-two-loop-amplitudes}.

\subsubsection{Two-loop self-energy diagram}
\label{sec:two-loop-real-virtual}

\begin{figure}[t]
\begin{center}
\includegraphics{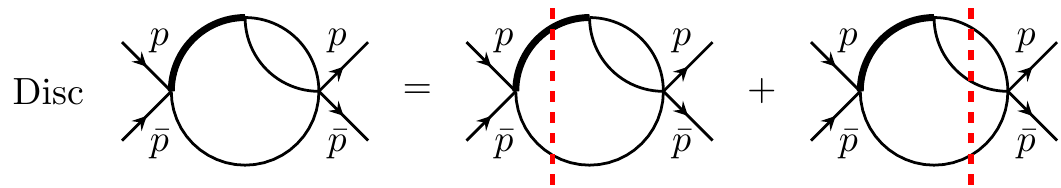}
\caption{Cutting equation for a two-loop self-energy diagram.}
\label{fig:two-loop-real-virtual-cutting-equation}
\end{center}
\end{figure}

In our first two-loop example we study a forward self-energy diagram, whose cutting equation is depicted in \cref{fig:two-loop-real-virtual-cutting-equation}. 
As illustrated in the figure, the forward diagram admits two cuts: a two- and a three-particle cut.
The two-particle cut is physical, but the three-particle cut is an unphysical $s$-channel cut which needs to be removed from the forward.
In this subsection we show how to compute the physical cut from the forward diagram and point out the differences with a direct calculation the cut diagram.
\newline

Let us start by computing the forward diagram, before proceeding to remove the unphysical cut in order to extract the moments of the physical cut.
The forward two-loop self-energy diagram is given by
\begin{align}
S = \big(\loopF Q^{2\eps} \big)^2 \, G_{1,1,1,1,0}~,
\label{eq:S_definition}
\end{align}
where $\loopF = (i \pi^{2-\eps} r_\Gam)^{-1}$ and $G_{1,1,1,1,0}$ is embedded in the integral topology
\begin{align}
G_{a_1,a_2,a_3,a_4,a_5} &=
\int \frac{\d^{4-2\eps}k ~ \d^{4-2\eps}\ell } {D_1^{a_1}\,D_2^{a_2}\,D_3^{a_3}\,D_4^{a_4}\,D_5^{a_5}} ~,
\label{eq:master_integral_B2a}
\end{align}
with denominators $D_i$ given by the following expressions in terms of $P=p+\pbar$,
\begin{align}
D_1 &= k^2-Q^2 ~, &
D_2 &= (k+P)^2 ~, &
D_3 &= \ell^2 ~, &
D_4 &= (\ell+k)^2 ~, &
D_5 &= (\ell+P)^2 ~.
\end{align}
We proceed to compute $G_{1,1,1,1,0}$ by establishing an appropriate differential equation. 
To this end, notice that raising the power of the massive propagator may be achieved by differentiation with respect to the mass $Q^2$, that is
\begin{align}
G_{2,1,1,1,0} = \frac{\d}{\d Q^2} G_{1,1,1,1,0} ~.
\label{eq:derivative_of_G11110}
\end{align}
Using IBP reduction, the integral on the left-hand side can be expressed in terms of simpler integrals
\begin{align}
G_{2,1,1,1,0} = 
-\frac{\left(Q^2 + (s-3 Q^2) \eps \right)G_{1,1,1,1,0}}{Q^2 \left(s-Q^2\right)}
-\frac{(2-3 \eps) G_{0,1,1,1,0}}{Q^2 \left(s-Q^2\right)}
+\frac{(1-\eps) G_{1,0,1,1,0}}{Q^2 \left(s-Q^2\right)} ~.
\label{eq:IBP_reduction_G21110}
\end{align}
The first integral on the right-hand side is the self-energy diagram
at hand (up to a prefactor), the integral on the left-hand side is its derivative, and the last two integrals on the right-hand side are simpler bubble-type integrals.
The latter can readily be computed exactly in $\eps$, producing
\begin{align}
G_{0,1,1,1,0} &= R_1(\eps) \, s^{1-2 \eps}  ~~~~~~~\text{with}~~
R_1(\eps) = -\frac{\pi^{4-2\eps}\,\Gam(1-\eps)^3 \Gam(-1+2\eps)}{\Gam(3-3 \eps)}\,e^{2 i \pi \eps} ~,
\label{eq:boundary_R1}
\\
G_{1,0,1,1,0} &= R_2(\eps) \, (Q^2)^{1-2\eps}  ~~~\text{with}~~
R_2(\eps) = \frac{\pi^{4-2\eps}\,\Gam(1-\eps)^2\Gam(\eps)\Gam(-1+2\eps)}{\Gam (2-\eps)} ~.
\label{eq:boundary_R2}
\end{align}
where we have performed the analytic continuation $(-s)^{-2\eps} \equiv (-s-i0)^{-2\eps} = e^{2 i \pi \eps} s^{-2\eps}$.
Inserting \cref{eq:derivative_of_G11110,eq:boundary_R1,eq:boundary_R2} into \cref{eq:IBP_reduction_G21110} thus produces a first-order linear differential equation for the integral $G_{1,1,1,1,0}$.
Exchanging the latter for $S$, see \cref{eq:S_definition}, 
yields the following differential equation
\begin{align}
\left( (1-\eps-\eps\,\w) - (1-\w) Q^2 \frac{\d}{\d Q^2} \right) S
&= 
-(2-3 \eps)\,\loopF^2 \,R_1(\eps)\,\w^{1-2 \eps} 
\nn
&\quad 
+(1-\eps)\,\loopF^2 \,R_2(\eps) ~.
\label{eq:diff_eq_S}
\end{align}
As we discussed in the previous subsection, we now insert a series ansatz for $S$ into this differential equation, to turn it into a difference equation for the series coefficients.
From the inhomogeneous terms in \cref{eq:diff_eq_S} one can infer that the forward diagrams will have the structure $S = f_1(\w) + \w^{-2\eps} f_2(\w)$, where $f_{1}(\w)$ and $f_{2}(\w)$ are regular functions of $\w$ close to the origin.
We thus proceed to make the series ansatz\footnote{
Alternatively, one may insert the general ansatz for two-loop diagrams in \cref{eq:two_loop_ansatz_for_forward} and derive that the corresponding coefficients $d_n$ all vanish.
}
\begin{align}
S = \sum_{n=0}^{\infty} c_n \, \omega^n 
+ \sum_{n=0}^{\infty} e_n \, \omega^{n-2\eps} ~.
\label{eq:S_ansatz}
\end{align}
Inserting this into \cref{eq:diff_eq_S} yields 
\begin{align}
0&=
\sum_{n=1}^{\infty} c_{n-1} (n-1+\eps) \omega^n 
- \sum_{n=0}^{\infty} c_n (n+1-\eps) \omega^n
\nn&
+ \sum_{n=1}^{\infty} e_{n-1} (n-1-\eps) \omega ^{n-2 \eps }
- \sum_{n=0}^{\infty} e_n (n+1-3\eps) \omega ^{n-2 \eps }
\nn&
-(2-3\eps) \loopF^2 R_1(\eps ) \omega ^{1-2 \eps }
+(1-\eps ) \loopF^2 R_2(\eps ) ~.
\end{align}
Equating same powers of $\omega$ produces two recursions, complete with boundary conditions:
\begin{align}
c_n &= \left(\frac{n-1+\eps}{n+1-\eps}\right) c_{n-1} ~~~\,\text{for}~n > 0 ~,~~
c_0 = \loopF^{2} R_2(\eps)~,
\\
e_n &= \left(\frac{n-1-\eps}{n+1-3\eps}\right) e_{n-1} ~~\text{for}~n > 1 ~,~~
e_1 = - \loopF^{2} R_1(\eps) ~,~~
e_0 = 0 ~.
\label{eq:coef_recursion}
\end{align}
The solutions to these elementary recursions are ratios of gamma functions,
\begin{align}
c_n &= \loopF^2 R_2(\eps) \,
\frac{\Gam(2-\eps)}{\Gam(\eps)} \,
\frac{\Gam(n+\eps)}{\Gam(n+2-\eps)}
~~~\text{for}~n \geq 0~,
\\
e_n &= - \loopF^2 R_1(\eps) \,
\frac{\Gam(3-3\eps)}{\Gam(1-\eps)} \,
\frac{\Gam(n-\eps)}{\Gam(n+2-3\eps)}
~~~\text{for}~n \geq 1~.
\label{eq:coef_solutions}
\end{align}
As a result, the forward self-energy diagram $S$ is now known as a series expansion around the origin:
\begin{align}
S = 
- \frac{\Gam(1-2\eps)^2\Gam(-1+2\eps)}{\Gam(1-\eps)^2\Gam(1+\eps)^2}
\left(\,
	\sum_{n=0}^{\infty} \frac{\Gam(n+\eps)}{\Gam (n+2-\eps)} \, \omega^n 
	+e^{2 i \pi  \eps } 
	\sum_{n=1}^{\infty} \frac{\Gam(n-\eps)}{\Gam(n+2-3\eps)} \omega^{n-2\eps} 
\right) ~.
\label{eq:S_solution}
\end{align}
These series can easily be recognised as representations of ${}_2F_1$-hypergeometric functions,
but for our purposes the current form is actually more useful.
Indeed, the aim of the remainder of this section is to extract
the Mellin moments of the physical cut in
\cref{fig:two-loop-real-virtual-cutting-equation} from the forward
amplitude diagram in \cref{eq:S_solution}.

Extracting the Mellin moments of the physical cut from the forward is done in the following way.
We construct a new function $\widetilde{S}$, which has the same branch cut discontinuity as $S$ around $\w \in [1,\infty)$, but does not possess a branch cut starting at the origin $\w = 0$.
In practice, we find such a function by means of the shifting procedure, as explained in \cref{sec:shifting-procedure}.
Applied to the series in \cref{eq:S_solution} this produces
\begin{align}
S \,\longrightarrow\, \widetilde{S} = - \left(1+e^{2 i \pi  \eps } \right)
\frac{\Gam(1-2\eps)^2\Gam(-1+2\eps)}{\Gam(1-\eps)^2\Gam(1+\eps)^2}
\sum_{n=1}^{\infty} \frac{\Gam(n+\eps)}{\Gam (n+2-\eps)} \, \omega^n ~.
\label{eq:S_shifted}
\end{align}
where we dropped an $\w$-independent term, without affecting the discontinuity.
The series coefficients of this new function $\widetilde{S}$ (in contrast to $S$) are well-defined. 
If we write $\widetilde{S} = \sum_{n=1}^{\infty} \widetilde{c}_n \w^n$, then its series coefficients $\widetilde{c}_n$ are equal to 
\begin{align}
\widetilde{c}_n = - \left(1+e^{2 i \pi  \eps } \right)
\frac{\Gam(1-2\eps)^2\Gam(-1+2\eps)}{\Gam(1-\eps)^2\Gam(1+\eps)^2}
\frac{\Gam(n+\eps)}{\Gam (n+2-\eps)} ~.
\label{eq:S_coefs}
\end{align}
Based on our arguments presented in \cref{sec:approach-two-loop-amplitudes} we claim that these series coefficients provide the Mellin moments of the physical cut on the right-hand side of the cutting equation in \cref{fig:two-loop-real-virtual-cutting-equation}.
The coefficients in \cref{eq:S_coefs} therefore constitute the main result of this example.
\newline

Let us now verify our claim.
To this end we shall compute the physical cut diagram explicitly.
One way to proceed is by applying reverse unitarity \cite{Anastasiou:2002yz} to the IBP reduction in \cref{eq:IBP_reduction_G21110}, in order to derive a differential equation for the cut diagram.
Alternatively, one may actually perform the phase-space integration directly.
In the latter approach one simply integrates a massless sub-bubble over a two-particle (one-mass) phase space.
The massless sub-bubble reads
\begin{align}
\mathrm{Bub} = i \pi^{2-\eps} \, \frac{\Gam(1-\eps)^2\,\Gam(\eps)}{\Gam(2-2\eps)} \, \frac{1}{(-k^2)^{\eps}} ~.
\label{eq:CutS_sub_bubble}
\end{align}
Because the massive line is cut, $k^2 = Q^2$, the bubble can be pulled out of the phase-space integral.
As a result, the cut diagram is given by
\begin{align}
\Cut_{\text{phys}} S 
&= 
2 \pi i \, e^{i \pi \eps} \, z^{\eps} (1-z)^{1-2\eps} 
\frac{\Gam(1-2\eps)^2\,\Gam(\eps)}{\Gam(2-2\eps)^2\,\Gam(1-\eps)\,\Gam(1+\eps)^2} ~.
\label{eq:CutS}
\end{align}
The Mellin moments can be computed exactly in $\eps$ due to the simple dependence on $z$. 
They are given by
\begin{align}
M_n\big[\!\Cut_{\text{phys}} S \big] &= 
2 \pi i \, e^{i \pi \eps}
\frac{\Gam(1-2\eps)^2\,\Gam(\eps)}{\Gam(2-2\eps)\,\Gam(1-\eps)\,\Gam(1+\eps)^2} 
\frac{\Gam(n+\eps)}{\Gam(n+2-\eps)}~.
\label{eq:CutS_moments}
\end{align}
Comparing this expression to the series coefficients in \cref{eq:S_coefs}, and making use of the identity 
$\Gam(1-2\eps)\Gam(1+2\eps) \left(1+e^{2 i \pi  \eps }\right) = 
2\Gam(1-\eps)\Gam(1+\eps)e^{i \pi  \eps }$, we can verify that
\begin{align}
\widetilde{c}_n & = \frac{1}{2\pi i} M_n\big[\!\Cut_{\text{phys}} S \big] ~.
\label{eq:CutS_claim}
\end{align}
This relation holds at all orders in $\eps$, as claimed.

\subsubsection{Two-loop crossed-box diagram}
\label{sec:two-loop-crossed-box}

\begin{figure}[t]
\begin{center}
\includegraphics{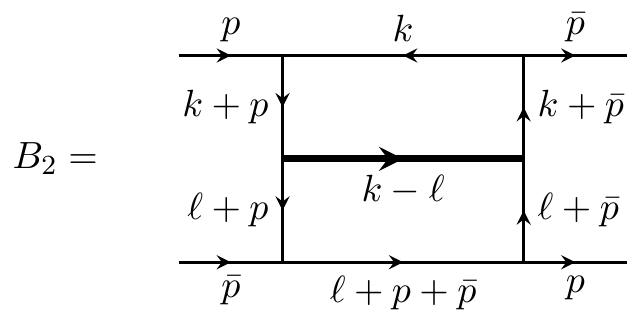}
\caption{The forward two-loop crossed-box diagram $B_2$.}
\label{fig:two-loop-crossed-box}
\end{center}
\end{figure}

We turn to our second example: the two-loop crossed-box diagram, depicted in \cref{fig:two-loop-crossed-box}.
This diagram is distinguished from previous examples in two key aspects.
First, it is sufficiently complicated so that it cannot be calculated
exactly in $\eps$, thereby providing a testing ground for the techniques of the previous subsection for computing the series coefficients of forward diagrams order-by-order in $\eps$ from differential equations.
Second, the diagram is the first example to admit a massive $u$-channel cut, for which a new procedure was developed also in the previous subsection.
In the example below we focus on these two aspects: first we compute the forward diagram, after which the moments of the physical cut are recovered by means of the shifting procedure and the replacement rules.
We finally cross-check our results against the literature.

Our first task is to compute the crossed-box diagram in \cref{fig:two-loop-crossed-box}.
It may be written as
\begin{align}
B_2 = \big(\loopF \, Q^{2\eps}\big)^2 \, s^3 ~ G_{1,1,1,1,1,1,1} ~,
\end{align}
where $G_{1,1,1,1,1,1,1}$ is one of the integrals in the following two-loop double-box topology
\begin{align}
G_{a_1,a_2,a_3,a_4,a_5,a_6,a_7} &=
\int \frac{\d^{4-2\eps}k ~ \d^{4-2\eps}\ell } {D_1^{a_1}\,D_2^{a_2}\,D_3^{a_3}\,D_4^{a_4}\,D_5^{a_5}\,D_6^{a_6}\,D_7^{a_7}} ~,
\label{eq:B2_topology}
\end{align}
where the denominators $D_i$ are given by 
\begin{align}
D_1 &= k^2 ~, &
D_2 &= (k+p)^2 ~, &
D_3 &= (k+\pbar)^2 ~, &
D_4 &= (\ell+p)^2 ~, \nn
D_5 &= (\ell+\pbar)^2 ~, &
D_6 &= (\ell+p+\pbar)^2 ~, &
D_7 &= (k-\ell)^2 - Q^2 ~.
\label{eq:B2_propagators}
\end{align}
There are fourteen master integrals for the topology in \cref{eq:B2_topology}. 
We use the following set of master integrals, as provided by \texttt{Litered} \cite{Lee:2012cn,Lee:2013mka},
\begin{align}
 M^B_1 \,&=\, G_{0,0,0,1,1,0,1} ~, &
 M^B_2 \,&=\, G_{0,0,1,0,0,1,1} ~, &
 M^B_3 \,&=\, G_{0,0,1,0,1,0,1} ~, \nn
 M^B_4 \,&=\, G_{0,0,1,1,0,0,1} ~, &
 M^B_5 \,&=\, G_{1,0,0,0,0,1,1} ~, &
 M^B_6 \,&=\, G_{0,0,2,1,0,0,1} ~, \nn 
 M^B_7 \,&=\, G_{2,0,0,0,0,1,1} ~, &
 M^B_8 \,&=\, G_{0,1,1,1,1,0,0} ~, \nn
 M_{9} \,&=\, G_{0,1,1,0,0,1,1} ~, &
 M_{10} \,&=\, G_{0,1,1,0,1,1,1} ~, &
 M_{11} \,&=\, G_{0,1,1,1,1,0,1} ~, \nn
 M_{12} \,&=\, G_{0,2,1,0,0,1,1} ~, &
 M_{13} \,&=\, G_{1,0,1,0,1,1,1} ~, &
 M_{14} \,&=\, G_{1,1,1,1,1,1,1} ~.
 \label{eq:B2_masters}
\end{align}
The integral of interest is the last (and most complicated) master integral $M_{14}$.
Following the notation introduced in \cref{sec:approach-two-loop-amplitudes}, the first eight integrals are marked with the superscript ``B'' to indicate that they can be readily computed as iterated bubble integrals.
These integrals are
\begin{align}
 M^B_1&=\frac{\pi ^{4-2 \eps } s^{1-2\eps} \w ^{-1+\eps} \Gam (1-\eps )^2 \Gam (-1+\eps) \Gam
   (\eps )}{\Gam (2-2 \eps )} ~,\nn
 M^B_2&=\frac{\pi ^{4-2 \eps } s^{1-2\eps} \w ^{-1+2 \eps} \Gam (1-\eps )^2 \Gam (\eps ) \Gam
   (-1+2 \eps)}{\Gam (2-\eps )} ~,\nn
 M^B_3&=\frac{\pi ^{4-2 \eps } s^{1-2\eps} \w ^{-1+2 \eps} \Gam (1-\eps )^2 \Gam (\eps ) \Gam
   (-1+2 \eps)}{\Gam (2-\eps )} ~,\nn
 M^B_4&=\frac{\pi ^{4-2 \eps } s^{1-2\eps} \w ^{-1+2 \eps} \Gam (1-\eps )^2 \Gam (\eps ) \Gam
   (-1+2 \eps) \, _2F_1(\eps ,-1+2 \eps;2-\eps ;-\w )}{\Gam (2-\eps )} ~,\nn
 M^B_5&=\frac{\pi ^{4-2 \eps } s^{1-2\eps} \w ^{-1+2 \eps} \Gam (1-\eps )^2 \Gam (\eps ) \Gam
   (-1+2 \eps ) \, _2F_1(\eps ,-1+2 \eps;2-\eps ;\w )}{\Gam (2-\eps )} ~,\nn
 M^B_6&=-\frac{\pi ^{4-2 \eps } s^{-2\eps} \w ^{2 \eps } \Gam (1-\eps ) \Gam (-\eps ) \Gam (2
   \eps ) \Gam (1+\eps) \, _2F_1(2 \eps ,1+\eps;2-\eps ;-\w )}{\Gam
   (2-\eps )} ~,\nn
 M^B_7&=-\frac{\pi ^{4-2 \eps } s^{-2\eps} \w ^{2 \eps } \Gam (1-\eps ) \Gam (-\eps ) \Gam (2
   \eps ) \Gam (1+\eps) \, _2F_1(2 \eps ,1+\eps;2-\eps ;\w )}{\Gam
   (2-\eps )} ~,\nn
 M^B_8&=-\frac{\pi ^{4-2 \eps} s^{-2\eps} \Gam (1-\eps )^4 \Gam (\eps )^2}{\Gam (2-2 \eps )^2} ~.
\label{eq:B2_boundary_integrals}
\end{align}
Being exact in $\eps$, these expressions are allowed to appear as inhomogeneous terms in differential equations for the six remaining unknown master integrals. 

We proceed to derive decoupled differential equations for the master integrals 
$M_9$ through $M_{14}$ of the form in \cref{eq:differential_equation_Mi}, using the Laporta reduction algorithm in \texttt{FIRE} \cite{Smirnov:2008iw,Smirnov:2013dia}.
Inserting the series ansatz \cref{eq:two_loop_ansatz_for_forward} for the two-loop forward master integrals,  the differential equations then transform into difference equations.
It turns out that three of those equations can be solved exact in $\eps$, producing series coefficients expressed as ratios of Gamma functions.
As a result, the ans\"atze are easily recognised as series representations of hypergeometric functions:
\begin{align}
M_{9}&=
\frac{\pi ^{4-2 \eps} s^{-2\eps} \w ^{2 \eps } \Gam (1-\eps )^2 \Gam(\eps ) \Gam (2 \eps) \, _3F_2(1,1,2 \eps ;2,2-\eps ;\w )}{\Gam (2-\eps)}
\nn&\quad
+\frac{\pi ^{4-2 \eps } s^{-2\eps} \w ^{\eps } \Gam (1-\eps )^2 \Gam (-1+\eps) \Gam(\eps ) \, _3F_2(1,1-\eps ,\eps ;2-2 \eps ,2-\eps ;\w )}{\Gam (2-2 \eps)} ~,
\nn
M_{12}&= 
- \frac{\pi ^{4-2 \eps } s^{-1-2\eps} \w ^{2 \eps } (1-3\eps+2\eps^2) \Gam(1-\eps )^2 \Gam (\eps ) \Gam (-1+2 \eps) \left(\, _3F_2(1,1,2 \eps ;2,1-\eps;\w )-1\right)}{\Gam(2-\eps)}
\nn&\quad 
+\frac{\pi ^{4-2 \eps } s^{-1-2\eps} \w ^{\eps} \Gam (1-\eps )^2 \Gam (-1+\eps) \Gam (\eps ) \, _3F_2(1,1-\eps ,\eps ;1-2\eps ,2-\eps ;\w )}{\Gam (1-2 \eps )} ~,
\nn
M_{13}&=
\frac{\pi ^{4-2 \eps } s^{-1-2\eps} \w ^{1+2 \eps} \Gam (-\eps )^2 \Gam (2+\eps) \Gam(1+2 \eps) \, _4F_3(1,1,2+\eps,1+2 \eps;2,2,2-\eps ;\w )}{\Gam (2-\eps )} ~.
\label{eq:B2_three_master_integrals}
\end{align}
These expressions are useful because the corresponding integrals may appear as inhomogeneous terms in differential equations for the remaining unknown integrals: $M_{10},M_{11}$ and $M_{14}$.

In the remainder we focus on the computation of $M_{14}$, which is the forward crossed-box diagram.
Inspecting the first-order differential equation for this integral
reveals that it is coupled to all other master integrals, in particular to the unknown integrals $M_{10}$ and $M_{11}$.
After decoupling those two, as described in \cref{sec:approach-two-loop-amplitudes}, we obtain a third-order differential equation for $M_{14}$.
As before we insert the series ansatz \cref{eq:two_loop_ansatz_for_forward} for the forward integral, which produces an eighth-order difference equation.
The latter can be solved order-by-order in $\eps$ in terms of harmonic numbers, c.f. \cref{eq:ansatz_series_coefficients}, using the strategy outlined in \cref{sec:extraction-of-series-coefs}.
If the full diagram $B_2 = (\loopF Q^{2\eps})^2 s^3 M_{14}$ is written as
\begin{align}
B_{2} = 
\sum_{n=1}^{\infty} c_n \, \omega^n 
+ \sum_{n=1}^{\infty} d_n \, \omega^{n-\eps}
+ \sum_{n=1}^{\infty} e_n \, \omega^{n-2\eps} ~,
\label{eq:B2_ansatz}
\end{align}
then its series coefficients are found to be
\begin{align}
c_n &= 
\frac{-2 S_{1,1}-4 S_{-2}-2 S_2}{\epsilon ^2}
+\frac{-16 S_{-2,1}-8 S_{1,-2}+4 S_{1,2}+4 S_{2,1}-18 S_{1,1,1}+20 S_{-3}+6 S_3}{\epsilon }
\nn&\quad
+80 S_{-3,1}+40 S_{-2,2}+40 S_{1,-3}
-8 S_{1,3}+8 S_{2,-2}-14 S_{2,2}-6 S_{3,1}-64 S_{-2,1,1}-56 S_{-4}
\nn&\quad
-32 S_{1,-2,1}-16 S_{1,1,-2}+52 S_{1,1,2}+52 S_{1,2,1}+48 S_{2,1,1}-110 S_{1,1,1,1}-14 S_4
- 8 \zeta_2 S_{1,1}
\nn&\quad
-16 \zeta_2 S_{-2}-8 \zeta_2 S_2 
+ \Ord(\eps)~,
\nonumber\\
d_n &=
-\frac{2 S_1}{\epsilon ^3}
+\frac{8 S_2-10 S_{1,1}}{\epsilon ^2}
+\frac{22 S_{1,2}+20 S_{2,1}-38 S_{1,1,1}-6 S_3-2 \zeta_2 S_1}{\epsilon }
-18 S_{1,3}-28 S_{2,2}
\nn&\quad
-14 S_{3,1}
+62 S_{1,1,2}+58 S_{1,2,1}+60 S_{2,1,1}-130 S_{1,1,1,1}+8 S_4 
-10 \zeta_2 S_{1,1}
+8 \zeta_2 S_2
\nn&\quad
-4 \zeta_3 S_1 
+ \Ord(\eps)~,
\nonumber\\
e_n & = 
-\frac{1}{\epsilon ^4}
-\frac{2 S_1}{\epsilon ^3}
-\frac{4 S_{1,1}+ 2 S_2}{\epsilon ^2}
-\frac{4 S_{1,2}+2S_{2,1}+8 S_{1,1,1}+2S_3}{\epsilon }
-4 S_{1,3}-2 S_{2,2}-2 S_{3,1}
\nn&\quad
-8 S_{1,1,2}-4 S_{1,2,1}-2 S_{2,1,1}-16 S_{1,1,1,1}-2 S_4 
+ \Ord(\eps)~,
\end{align}
where $S_{\vec{\ell}} \equiv S_{\vec{\ell}\,}(n-1)$.
We have checked the validity of the representation \cref{eq:B2_ansatz} for the forward crossed-box diagram, by reconstructing from the infinite sums the full diagram in terms of harmonic polylogarithms and performing a numerical cross-check using \texttt{SecDec} \cite{Carter:2010hi,Borowka:2012yc,Borowka:2015mxa}. 
These series coefficients now form the starting point for the next phase, which is to extract the Mellin moments of the corresponding physical $s$-channel cut diagram.

\begin{figure}[t]
\includegraphics{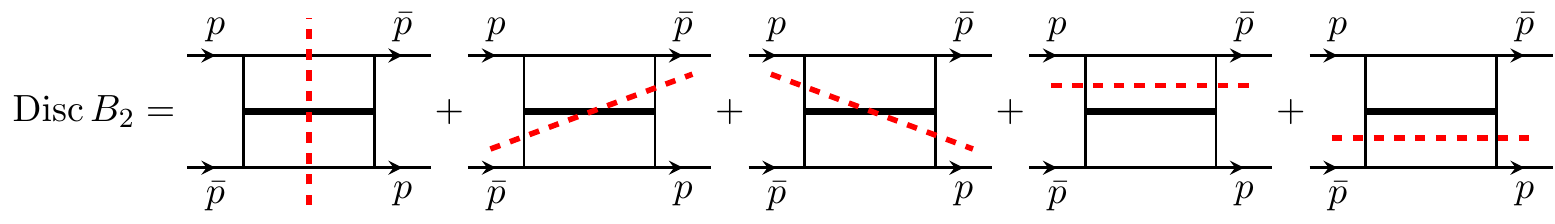}\\
\vspace{-0.25cm}\\
\includegraphics{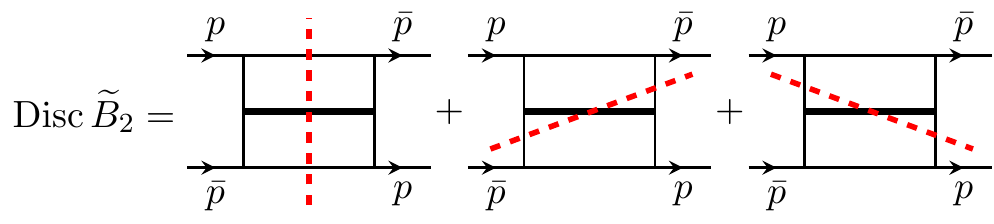}\\
\vspace{-0.25cm}\\
\includegraphics{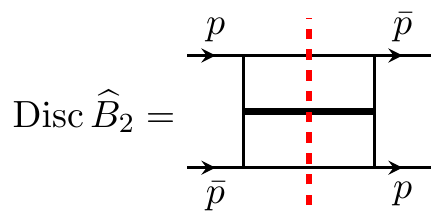}
\caption{Cutting equation for the forward diagram $B_2$ (top line), for the modified forward diagram $\widetilde{B}_2$ which does not contain the unphysical branch cut along $\w \in (-\infty,0]$ (middle line), and for $\widehat{B}_2$ which does not contain the other unphysical branch cut along $\w \in (-\infty,-1]$ either (bottom line). The series coefficient of $\widehat{B}_2$ are equal to the Mellin moments of the physical cut.
}
\label{fig:B2_cutting_equations}
\end{figure}

Inspecting the analytical structure of $B_2$ from its representation in terms of harmonic polylogarithms reveals three branch cuts.
They are located along the real axis in the domains $\w \in (-\infty,0]$, $\w \in (-\infty,-1]$ and $\w \in [1,\infty)$,
which correspond to massless $u$-channel cuts, massive $u$-channel cuts and the massive $s$-channel cut, respectively.
The first and second types of branch cuts in the forward diagram are unphysical; they will be removed by performing the shifting procedure and applying replacement rules, respectively.
Let us start with the shifting procedure.
As we have seen in previous examples, this amount to the transformation
\begin{align}
B_{2} ~\longrightarrow~ \widetilde{B}_{2} 
&\,=\, 
\sum_{n=1}^{\infty} (c_n +d_{n+\eps} + e_{n+2\eps}) \,\omega^{n} 
\,\equiv\, 
\sum_{n=1}^{\infty} \widetilde{c}_n \,\omega^{n} ~.
\label{eq:B2_tilde}
\end{align}
More explicitly, the newly defined coefficients $\widetilde{c}_n$ are given by
\begin{align}
\widetilde{c}_n &= 
-\frac{1}{\epsilon ^4}
-\frac{4 S_1}{\epsilon ^3}
+\frac{
10 S_2
-16 S_{1,1}
-4 S_{-2}
-6 \zeta _2
}{\epsilon ^2}
+\frac{
-20 S_3
+40 S_{1,2}
+40 S_{2,1}
-64 S_{1,1,1}
}{\epsilon }
\nn&\quad
+\frac{
20 S_{-3}
-16 S_{-2,1}
-8 S_{1,-2}
-20 \zeta _2 S_1
}{\epsilon }
+40 S_4
-100 S_{2,2}
-80 S_{3,1}
+160 S_{1,1,2}
\nn&\quad
+160 S_{1,2,1}
+160 S_{2,1,1}
-256 S_{1,1,1,1}
-56 S_{-4}
+80 S_{-3,1}
+40 S_{1,-3}
-84 S_{1,3}
\nn&\quad
+40 S_{-2,2}
+8 S_{2,-2}
-64 S_{-2,1,1}
-32 S_{1,-2,1}
-16 S_{1,1,-2}
-16 \zeta _2 S_{-2}
+44 \zeta _2 S_2
\nn&\quad
-72 \zeta _2 S_{1,1}
-4 \zeta _3 S_1
-55 \zeta _4
+\Ord(\eps) ~.
\end{align}
where again $S_{\vec{\ell}} \equiv S_{\vec{\ell}\,}(n-1)$.
In order to arrive at this form for the series coefficients $\widetilde{c}_n$, we have made use of the package \texttt{HarmonicSums} to expand the harmonic numbers $S_{\vec{\ell}\,}(n+k\eps)$, which appear in the coefficients $d_{n+\eps}$ and $e_{n+2\eps}$ as a result of shifting, as a Taylor series in $\eps$.
From the formula in \cref{eq:B2_tilde} it is clear that $\widetilde{B}_2$ is regular at the origin, so we have successfully removed the branch cut along $\w \in (-\infty,0]$ from the forward diagram.
Crucially, the discontinuities around the remaining two branch cuts are unchanged.
This can be verified by explicitly computing and comparing the discontinuity of both $B_2$ and $\widetilde{B}_2$ 
using the \texttt{HPL} package \cite{Maitre:2005uu,Maitre:2007kp}\footnote{Specifically, using the function \texttt{HPLAnalyticContinuation} in that package.}.
In terms of cutting equations, this elimination of unphysical branch cut in the forward diagram is to be interpreted as the elimination of cut diagrams on the right-hand side of the cutting equation, as indicated by the first two lines in \cref{fig:B2_cutting_equations}.

In the next stage we modify the forward diagram even further, in such a way that the second unphysical branch cut is removed as well.
At the level of individual harmonic polylogarithms this task is performed simply along the lines of the example in \cref{sec:approach-two-loop-amplitudes}.
The results translate to replacement rules for the harmonic numbers. 
In particular, harmonic numbers with only positive indices do not need to be altered: the corresponding ``resummed functions'' do not contain unphysical branch cuts.
The first two orders in $\eps$ of $\widetilde{c}_n$ therefore do not need to be modified.
For the remaining harmonic numbers we apply the following replacement rules.
We recall that these rules are derived in a diagram-independent way. 
At order $\eps^{-2}$ we need a single replacement rule:
\begin{align}
S_{-2} \rightarrow - \tfrac{1}{2} \zeta_2 ~.
\end{align}
Replacement rules at order $\eps^{-1}$ are:
\begin{align}
S_{-3} &\rightarrow - \tfrac{3}{4} \zeta_3 ~, \nonumber \\
S_{-2,1} &\rightarrow - \tfrac{5}{8} \zeta_3 ~, \nonumber \\
S_{1,-2} &\rightarrow - \tfrac{1}{8} \zeta_3 - \tfrac{1}{2} \zeta_2 S_1 ~.
\end{align}
And finally, replacement rules at order $\eps^0$ are given by:
\begin{align}
S_{-4} &\rightarrow 
-\tfrac{7}{8} \zeta_4 
~,\nn
S_{-3,1} &\rightarrow 
-\tfrac{11}{4}\zeta _4
-\tfrac{1}{2}\zeta_2 \log^2{2}
+\tfrac{7}{4} \zeta_3 \log{2}
+2\mathrm{Li}_4\!\left(\tfrac{1}{2}\right)
+\tfrac{1}{12} \log^4{2}
~, \nn
S_{1,-3} &\rightarrow 
+\tfrac{15}{8}\zeta _4
+\tfrac{1}{2}\zeta_2 \log^2{2}
-\tfrac{7}{4} \zeta_3 \log{2}
-2\mathrm{Li}_4\!\left(\tfrac{1}{2}\right)
-\tfrac{1}{12} \log^4{2}
-\tfrac{3}{4} \zeta_3 S_1
~, \nn
S_{-2,2} &\rightarrow 
+\tfrac{51}{16} \zeta_4
+\zeta_2 \log^2{2}
-\tfrac{7}{2} \zeta_3 \log{2}
-4 \mathrm{Li}_4\!\left(\tfrac{1}{2}\right)
-\tfrac{1}{6} \log^4{2}
~, \nn
S_{2,-2} &\rightarrow 
-\tfrac{65}{16} \zeta_4
-\zeta_2 \log^2{2}
+\tfrac{7}{2} \zeta_3 \log{2}
+4 \mathrm{Li}_4\!\left(\tfrac{1}{2}\right)
+\tfrac{1}{6} \log^4{2}
-\tfrac{1}{2} \zeta_2 S_2
~, \nn
S_{-2,1,1} &\rightarrow 
+\tfrac{5}{16} \zeta_4
+\tfrac{1}{4} \zeta_2 \log ^2{2}
-\tfrac{7}{8} \zeta_3 \log{2}
-\mathrm{Li}_4\!\left(\tfrac{1}{2}\right)
-\tfrac{1}{24} \log^4{2}
~, \nn
S_{1,-2,1} &\rightarrow 
-\tfrac{3}{16} \zeta_4
-\tfrac{5}{8} \zeta_3 S_1
~, \nn
S_{1,1,-2} &\rightarrow 
-\tfrac{13}{8} \zeta_4
-\tfrac{1}{4} \zeta_2 \log^2{2}
+\tfrac{7}{8} \zeta_3 \log{2}
+\mathrm{Li}_4\!\left(\tfrac{1}{2}\right)
+\tfrac{1}{24}\log ^4{2}
-\tfrac{1}{8} \zeta _3 S_1
-\tfrac{1}{2} \zeta_2 S_{1,1}
~.
\end{align}
After making these replacements the series coefficients $\widetilde{c}_n$ become $\widehat{c}_n$, given by
\begin{align}
\widehat{c}_n &= 
-\frac{1}{\epsilon ^4}
-\frac{4 S_1}{\epsilon ^3}
+\frac{10 S_2-16 S_{1,1}-4 \zeta _2}{\epsilon ^2}
\nn&\quad
+\frac{-20 S_3+40 S_{1,2}+40 S_{2,1}-64 S_{1,1,1}-16 \zeta _2 S_1-4\zeta_3}{\epsilon }
\nn&\quad
+40 S_4
-84 S_{1,3}
-100 S_{2,2}
-80 S_{3,1}
+160 S_{1,1,2}
+160 S_{1,2,1}
\nn&\quad
+160 S_{2,1,1}
-256 S_{1,1,1,1}
+40 \zeta _2 S_2
-64 \zeta _2 S_{1,1}
-12 \zeta _3 S_1
-24 \zeta _4
+\Ord(\eps) ~.
\label{eq:coefficients_c_hat}
\end{align}
These coefficients $\widehat{c}_n$ constitute the main result of this example section.
We have checked explicitly that resumming these coefficients yields an expression 
\begin{align}
\widehat{B}_{2} = \sum_{n=1}^{\infty} \widehat{c}_n \omega^n~,
\label{eq:B2_hat}
\end{align}
which only has a branch cut along $\w \in [1,\infty)$ and whose discontinuity along that branch cut is the same as for the original diagram $B_{2}$.
This means that these series coefficients $\widehat{c}_n$ in \cref{eq:coefficients_c_hat} must be equal to the Mellin moments of the sum of physical cuts of the forward diagram $B_2$!
We claim that
\begin{align}
\frac{1}{2\pi i}{\mathcal M}_n[\Cut_{\text{phys}} B_{2}] = \widehat{c}_n ~.
\label{eq:B2_moments_from_series_coefs}
\end{align}

\begin{figure}[t]
\begin{center}
\includegraphics{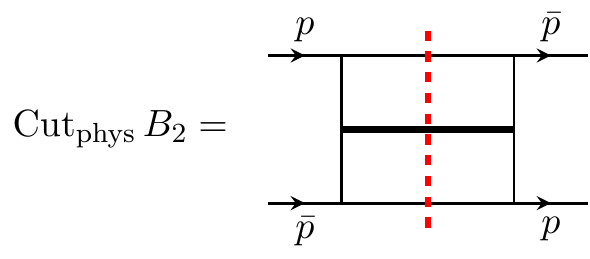}
\caption{The only physical cut of the two-loop forward crossed-box diagram.}
\label{fig:two-loop-crossed-box-physical-cut}
\end{center}
\end{figure}

The validity of the above statement can be verified by comparing the coefficients $\widehat{c}_n$ against an explicit computation of the Mellin moments of the physical cut of the forward diagram $B_2$, depicted in \cref{fig:two-loop-crossed-box-physical-cut}.
An explicit result for this particular cut diagram was given in eq.~(B.21) of \citeR{Anastasiou:2002yz}.
Correcting for small misprints (see appendix A of \citeR{Pak:2011hs}) and adopting our normalisation convention, we have that
\begin{align}
\Cut_{\text{phys}} B_{2} &= 
- 2 \pi i \, \mathcal{N}(\eps)\, z^{2\eps} (1-z)^{-1-4\eps} \bigg( \!
	- \frac{4}{\eps^3} 
	+ \frac{16}{\eps^2} 
	+ \frac{z(z-8)-89}{6\eps} 
	\nn&\hspace{7mm}
	+ \tfrac{1}{9}(z-1)(1+2z)
	+\tfrac{2}{27}(z-1)(13z-16) \eps
	+\Ord(\eps^2)
\bigg)
\nn&\hspace{4mm}
- 2 \pi i \, \mathcal{N}(\eps)\, z^{2\eps} (1-z)^{-1-2\eps} \bigg( \!
	- \frac{2 \log z}{\eps^2} 
	+ \frac{1}{\eps} \Big[ 
		2 \log^2 z 
		\nn&\hspace{7mm}
		+ \log z \, \big(4\log(1-z)+8\big) 
		+\tfrac{1}{6}(1-z)(z-7) 
		\Big]
	- \tfrac{4}{3} \log^3 z
	\nn&\hspace{7mm}
	- \log^2 z \,\big(8+2\log(1-z)\big)
	+36 \big(\Trilog(z)-\zeta_3\big)
	+\tfrac{1}{9}(z-2z^2+1)
	\nn&\hspace{7mm}
	- \log z \,\big(20\zeta_2+16\Dilog(z)+16\log(1-z)+4\log^2(1-z)+8\big)
	\nn&\hspace{7mm}
	+\tfrac{1}{3}(z-1)(z-7)\log(1-z) 
	+ \Ord(\eps)
\bigg) ~.
\label{eq:B2_cut}
\end{align}
where $\mathcal{N}(\eps) = \frac{\Gam(1-2\eps)^4}{\Gam(1-4\eps)\,\Gam(2-2\eps)^2\,\Gam(1-\eps)^2\,\Gam(1+\eps)^2}$.
After expanding the factors $(1-z)^{-1-k\eps}$ in terms of plus-distributions and taking the Mellin transform of this equation, we find perfect agreement with our formula in \cref{eq:B2_moments_from_series_coefs}, which expresses the Mellin moments in terms of series coefficients of the (modified) forward diagram.
This concludes the example.

\section{Conclusions}
\label{sec:conclusions}

In this paper we have presented a method for computing Mellin moments of single-particle inclusive cross sections, such as Drell-Yan and Higgs production, directly from forward scattering diagrams by invoking unitarity in the form of cutting equations. 
Due to the non-inclusive nature of these processes, the cutting equations contain unphysical cuts. 
The main achievement of this paper is to a provide diagram-independent prescription for ``removing'' such unphysical cut contributions to the discontinuity of a forward diagram, once these are expressed in the reciprocal $\omega=1/z$ variable.
The removal occurs through a complex shift in the summation index, and through a replacement rule dictionary for harmonic sums in the results.
After this, the modified sum over powers of $\w$ reproduces precisely the desired sum of physical cuts, and the coefficients are precisely the Mellin moments of the corresponding contribution to the cross section. 
We have demonstrated our method for various one- and two-loop diagrams.

The approach of this paper is conceptually similar to the computation of three-loop DIS splitting functions \cite{Vogt:2004mw,Moch:2004pa}. 
While DIS is a fully-inclusive process, our method provides a non-trivial extension to semi-inclusive processes.  Other methods exist for obtaining cross sections of semi-inclusive processes, but they do not make use of the optical theorem or cutting equations.
For example, one very successful approach \cite{Anastasiou:2002yz} computes cut diagrams as solutions to differential equations. 
Technically the latter need to be augmented with boundary conditions coming from a separate calculation (typically expansion-by-regions). 
In our approach the boundary conditions to difference equations for the Mellin moments are provided by the results for bubble-type loop integrals. 
In these other approaches calculations are moreover performed in $z$-space, except in \cite{Mitov:2005ps}.

Our method thus provides a new means of computing semi-inclusive cross sections, at least up to two-loop order. 
Since the main ingredients to the method are forward loop diagrams, as opposed to cut diagrams, it is particularly useful as an alternative to corrections involving real radiation, but provides no alternative way to compute virtual corrections. 
Being exclusively made out virtual diagrams, numerical cross-checks may be performed in a uniform way for all contributions (see the two-loop examples in the previous section). 

In regards to extending our method beyond two-loop order, we note that the work in this paper is based on an analysis of (un)physical branch cuts and the assumption that the solution space is spanned by harmonic sums. Both aspects will need to be reviewed at higher loops, but we are hopeful that progress can be made towards single-scale cross sections at N$^3$LO.

\paragraph{Acknowledgments}
We gratefully acknowledge Sven Moch, Jos Vermaseren and Andreas Vogt
for helpful discussions, as well as Mert Aybat and Lisa Hartgring for
early collaboration.  This work was supported by the Research
Executive Agency (REA) of the European Union under the Grant
Agreements number PITN-GA-2010-264564 (LHCPhenoNet) and
PITN-GA-2012-316704 (HIGGSTOOLS); by the Foundation for Fundamental
Research of Matter (FOM), programme 156, ``Higgs as Probe and Portal''; by the Dutch National Organization
for Scientific Research (NWO).

\bibliography{bibliography}
\end{document}